\documentclass[11pt,a4paper]{article}
\usepackage{jheppub}
\usepackage{lipsum}
\usepackage{braket}
\usepackage{tensor}
\usepackage{dsfont}
\usepackage{tcolorbox}
\usepackage{tikz}
\usetikzlibrary{decorations.pathreplacing,calligraphy}
\usetikzlibrary{decorations.pathreplacing}
\usetikzlibrary{decorations.pathmorphing, patterns,shapes}
\usepackage{bm}

\newcommand{\cm}{\mathcal{M}}
\newcommand{\co}{\mathcal{O}}

\newcommand{\cO}[1]{\mathcal{O}_{#1}(x_{#1})}
\newcommand{\phu}{\varphi}
\newcommand{\op}[1]{\boldsymbol{#1}}
\title{Broken (super) conformal Ward identities at finite temperature}

\author[a]{Enrico Marchetto,}
\author[b]{Alessio Miscioscia}
\author[b]{and Elli Pomoni}
\affiliation[a]{Mathematical Institute, University of Oxford, \\
	Andrew Wiles Building, Woodstock Road, Oxford, OX2 6GG, UK}
\affiliation[b]{Deutsches Elektronen-Synchrotron DESY, Notkestr. 85, 22607 Hamburg, Germany}
\emailAdd{enrico.marchetto@maths.ox.ac.uk}
\emailAdd{alessio.miscioscia@desy.de}
\emailAdd{elli.pomoni@desy.de}
\preprint{DESY-23-083}
\abstract{
When a (super) conformal field theory is placed on a non-trivial manifold, the (super) conformal symmetry is broken. However, it is still possible to derive broken Ward identities for these broken symmetries, which provide additional constraints on the theory.
We derive and apply the broken Ward identities associated with the (super) conformal group on the thermal manifold  $\mathcal{M}_\beta = S_\beta^1 \times \mathbb{R}^{d-1}$ and $\mathcal{M} = T^2 \times \mathbb{R}^{d-2}$.
The novel constraints not only systematically reproduce known results, including an implicit formulation of the generalized Cardy formula, but also elegantly relate the thermal energy spectrum with the conformal spectrum.}

\begin{document}
	\maketitle
	\flushbottom
	\section{Introduction and Summary} \label{sec: Introduction} 
 While there have been remarkable strides in our understanding of non-perturbative effects in Quantum Field Theory (QFT), our current knowledge of QFT at finite temperature remains relatively limited. In the space of QFTs, a special role is played by Conformal Field Theories (CFTs), not only because they correspond to the fixed points of the renormalization group flows, but also because conformal invariance implies powerful constraints on the CFT data studied by the conformal bootstrap program. 
Turning on a temperature $T$ can be understood as placing the CFT on the thermal manifold $\mathcal M_\beta = S_\beta^1\times \mathbb R^{d-1}$, where $\beta = 1/T$ is the length of the circle. 
The thermal bootstrap, initiated by \cite{Iliesiu:2018fao,Iliesiu:2018zlz}, and its relations to black hole physics (using the  AdS/CFT correspondence) \cite{Witten:1998zw,Horowitz:1999jd,deHaro:2000vlm,Son:2002sd,Aharony:2003sx,Casalderrey-Solana:2011dxg,Ammon:2015wua,Nastase:2015wjb,Caron-Huot:2022lff,Dodelson:2023vrw,Dodelson:2022yvn}, is currently a very active field of research to which we wish to contribute with this paper.

In QFT (continuous) symmetries are associated with differential equations correlation functions must obey, known as Ward identities. When a symmetry is broken we do not expect the corresponding Ward identity to hold, but we expect it will be corrected by an additional \textit{breaking term}. In this paper, we provide a general way to compute the breaking term for any continuous symmetry, with particular interest on the Conformal Group, on a generic manifold $\mathcal M$. Focusing on the thermal manifold $\mathcal M_\beta$, dilatations, special conformal transformations, and rotations including the time direction (the compactified direction) are broken, while translations and rotations including only space directions are preserved. 
This work is devoted to the computation of the breaking terms for the broken Ward identities.
Since the currents associated with the Conformal Group generators are proportional to the stress-energy tensor we show that every broken Ward identity is a non-trivial identity between an $n$-point function and an integrated $(n+1)$-point function including (some components of) the stress-energy tensor. We derive systematically those equations and we start exploring some of their physical implications. They have to be satisfied for every local CFT at finite temperature and provide constraints which go beyond the familiar zero-temperature conformal bootstrap conditions. We further show that some of them are related to the Hamiltonian and the momentum on the thermal manifold $\mathcal M_\beta$. In particular, the Hamiltonian is related to the dilatation operator while the momentum is related to the boost operator. Those relations are derived and discussed for the first time in this paper for general dimension.
Notably, the thermal energy spectrum is directly related to the zero-temperature spectrum.

What is more, supersymmetry is expected to be broken at finite temperature due to the anti-periodicity of fermions on the thermal circle $S_\beta^1$. We therefore extend the study of broken and unbroken Ward identities to the case of superconformal field theories, showing that the R-symmetry is preserved while supersymmetry and superconformal transformations are broken but are associated with equations that compare an $n$-point function with an $(n+1)$-point function including the supercurrent $G_{\alpha}^{\mu I}$.

 Because the geometry of the manifold $\mathcal M_\beta$ is conformally flat, the two-point functions can be expanded in an operator product expansion (OPE), which however will converge only in a suitable region  \cite{Iliesiu:2018fao, Gobeil:2018fzy}. Exploring broken Ward identities  in this regime, we check their consistency  with the results of \cite{Iliesiu:2018fao}. The broken Ward identities have to be satisfied for any correlation function at finite temperature and therefore could be used to enrich the bootstrap problem for correlation functions at finite temperature \cite{Alday:2020eua,Dodelson:2023vrw,Caron-Huot:2008vbk,Dodelson:2022yvn}. 
 As a preliminary exercise in this direction, 
we show (in Appendix \ref{sec: Exact computation of phiphi for a free scalar theory}) that the bootstrap problem can be exactly solved in the case of a free scalar field in four dimensions.

 Finally, in order to show the generality of our methods, we compute the same broken Ward identities for another interesting case, namely for the manifold $\mathcal M = T^2 \times \mathbb R^{d-2}$. This manifold is another interesting example of a non-trivial manifold in which the CFT can be placed. This geometry also attracted recent interest for different reasons \cite{Benjamin:2023qsc,Luo:2022tqy}.

 The outline of the paper is the following: \begin{itemize}
     \item[$\star$] In Section \ref{sec: Broken Ward Identities at Finite Temperature} the derivation of the broken Ward identities on a general manifold is presented. We apply this method to the thermal case.
     \item[$\star$] In Section \ref{sec: Conformal group} we apply the identities computed above for the case of the Conformal Group. We show that from those identities it is possible to read the operatorial expression of the Hamiltonian and the momentum at finite temperature (see Subsection \ref{ssec: ham and mom}). Finally, we combine broken Ward identities and operator product expansion and we show the consistency of broken Ward identities with known results.
      \item[$\star$] In Section \ref{sec: Superconformal BWI} we enlarge the study to the case of superconformal quantum field theories. We produce broken Ward identities and we show that R-symmetry is preserved. We also discuss the $\mathrm U(1)_Y$ bonus symmetry in $\mathcal N = 4$ SYM at finite temperature and the corresponding non-renormalization theorem.
     \item[$\star$] In Section \ref{appendix:TorusCase} we repeat the discussion of broken Ward identities for the case of the manifold $\mathcal M = T^2 \times \mathbb R^{d-2}$ producing broken Ward identities for this geometry.
 \end{itemize}

A number of technical points and derivations are presented in the appendices. In particular more comments on the dilatation broken Ward identity are presented in Appendix \ref{Appendix1}. Some explicit checks of the equations derived in the paper can be found in Appendix \ref{appendix A}. The solution for the free scalar theory in four dimensions obtained directly by solving the bootstrap problem is provided in Appendix \ref{sec: Free Scalar Tests of the Broken Ward Identities}. Finally some general comments on the broken Ward identities when combined with the operator product expansion can be read in Appendix \ref{Susy in OPE}.
 
	\section{Broken Ward identities at finite temperature} \label{sec: Broken Ward Identities at Finite Temperature}
    	It is known that one-, two- and three-point functions of a generic CFT living on the flat space-time are completely fixed, up to constants, by the symmetries composing the Conformal Group. These symmetries include translations, dilatations, rotations, and special conformal transformations. At the quantum level, each of these symmetries is associated with a Ward identity, which represents a specific differential equation that correlation functions must satisfy. In the flat spacetime $\mathbb{R}^d$, these differential equations are easily solvable and they return known results \cite{Osborn:1993cr, Rychkov:2016iqz,Simmons-Duffin:2016gjk,Gopakumar:2022kof}. 
     
    In this Section, we extract a Ward identity (or a set of Ward identities) for each symmetry transformation on the thermal manifold $\mathcal M_\beta = S_\beta^1\times \mathbb R^{d-1}$. We do not expect the symmetries to be preserved in general. For instance, we expect dilatations to be broken because of the insertion of a scale in the theory, represented by the temperature $T$ (in the following appearing trough its inverse $\beta = 1/T$\footnote{In the following we will use the units in which the Boltzmann constant is set to $k_B = 1$.}). In the derivation procedure, the non-triviality of the geometry leads to a modification of the standard flat space Ward identity: we will interpret this modification as a \emph{breaking} of the Ward identity. Although the primary focus of this paper is the thermal manifold, this Section can be extended to the case of a theory on a generic manifold $\mathcal M$. Therefore, we present the initial steps in this Section within a general framework. In Section \ref{appendix:TorusCase} we will study the case of a (super) conformal field theory on the manifold $\mathcal M = T^2\times \mathbb R^{d-2}$. 
  \newline Let us consider a generic CFT on $\mathbb R^{d}$. The observables of this theory can be constructed out of $n$-point correlation functions
  \begin{equation}
      \Braket{\cO{1}\dots \cO{n}} \ .
  \end{equation}
  These observables, as a consequence of the invariance under the global Conformal Group, satisfy specific properties. In this work, we are interested in studying how such properties are modified when the same correlation function is computed on a non-trivial manifold $\mathcal M$, where the conformal group is in general expected to be broken,\footnote{This can be understood as the breaking of some of the covariantly constant conformal Killing vectors on a non conformally equivalent to $\mathbb R^d$ manifold. For the purposes of this paper we are interested in specific manifolds with some non-broken symmetries, for example $S_1\times \mathbb R^{d-1}$ or $T^2\times \mathbb R^{d-2}$.}
	\begin{equation}
		\Braket{\cO{1}\dots \cO{n}}_{\mathcal M} \ , \label{eq: ncorrfunct}
	\end{equation}
	where $\braket{\dots}_{\mathcal M}$ indicates the correlation function on the manifold $\mathcal M$, and where it is understood that the operators satisfy specific boundary conditions intrinsically tied to the features of $\mathcal{M}$\footnote{As a clarifying example, in the bulk of the paper we will consider operators on the thermal manifold $\mathcal{M}_{\beta}=S^{1}_{\beta} \times \mathbb{R}^{d-1}$: the geometrical features require each operator to be periodic or antiperiodic over the thermal circle $S^{1}_{\beta}$.}. 
 
 The derivation of zero temperature Ward identities, i.e. Ward identities in the ambient flat spacetime manifold, is known in the literature (see for example \cite{DiFrancesco:1997nk,Papadodimas2010}). We start from a theory described by a formal action $S$ and we define an infinitesimal symmetry transformation $\varphi' (x) = \varphi(x) -i \omega_a \op G_a \varphi(x)$, where 
    \begin{equation}
	   i \omega_{a}\op G_a \varphi(x) = \omega_{a}\frac{\delta x^\mu}{\delta \omega_a}\partial_\mu \varphi - \omega_{a}\frac{\delta \mathcal F}{\delta \omega_a} \ .   \label{eq: symmtrans} 
	\end{equation}
 Here $\mathcal F$ is a function that parameterizes the variation of the field $\varphi$ under the symmetry generated by $\op G_a$.
 We will focus in Section \ref{sec: Conformal group} on the Conformal Group for which the action of the generators $\op G_{a}$ on a generic correlator are known and listed in Table \ref{tab: currents and generators}. We extend in Section \ref{sec: Superconformal BWI} the results for the superconformal case.
\newline  The correlation function \eqref{eq: ncorrfunct} can be represented as a path integration over the fundamental fields 
    \begin{equation}
         \Braket{\cO{1}\dots \cO{n}}_{\mathcal M} = \frac{1}{\mathcal Z}\int [d \phu] \  \mathcal O_1(x_1) \ldots \mathcal O_n(x_n) e^{-S[\varphi]} \ ;
    \end{equation}
    since the fundamental fields $\phu$ are integration variables, we can apply the transformation \eqref{eq: symmtrans} as a functional change of variables
    \begin{equation}\label{eq: Correlator identity}
         \Braket{\cO{1}\dots \cO{n}}_{\mathcal M} = \frac{1}{\mathcal Z}\int [d \phu']\ \left (\mathcal O_1(x_1)+\delta \mathcal O_1(x_1)\right) \ldots \left (\mathcal O_n(x_n)+\delta \mathcal O_n(x_n) \right) e^{-S'[\varphi']} \ ,
    \end{equation}
    where $\delta \mathcal{O}_{i}(x_i)$ denotes the effect of the transformation \eqref{eq: symmtrans} on the $i$-th local operator.
    If the theory is assumed to be invariant under rigid symmetry transformations, the transformed action on a generic manifold $\mathcal M$ can be written as \begin{equation} \label{eq: S transf}
        S'[\varphi'] =  S[\varphi]+ \int_{\mathcal M}d^d x  \sqrt{g} \ 
        \nabla_\mu^{\phantom{\mu}} \omega_a^{\phantom{\mu}}(x) J_a^{\mu}(x)  \ ,
    \end{equation}
    where $g$ is the metric on the manifold $\mathcal M$, and $J^{\mu}_{a}$ are the current operators associated with the conservation of the symmetry generated by $\op G_a$. In a Lagrangian theory, the current operators admit an explicit realization where the canonical stress-energy tensor appears
    \begin{equation}
        J_a^\mu =\frac{\delta x^\mu}{\delta \omega_a}\mathcal L - \frac{\partial \mathcal L}{\partial \partial _\mu \varphi}\partial_\nu \varphi \frac{\delta x^\nu }{\delta \omega_a} + \sum_{\phu}\frac{\partial \mathcal L}{\partial \partial_\mu \varphi}\frac{\delta \mathcal F}{\delta \omega_a}= T_{\text{can.}}^{\mu \nu}\frac{\delta x_\nu}{\delta \omega_a} + \sum_{\phu}\frac{\partial \mathcal L}{\partial \partial_\mu \varphi}\frac{\delta \mathcal F}{\delta \omega_a}\ .
    \end{equation}
   By focusing solely on the currents associated with the symmetries of the Conformal Group, we can reshape them in terms of the symmetric, traceless stress-energy tensor $T^{\mu \nu}$\footnote{The current associated with spacetime translations is $a_{\nu}T_\text{can}^{\mu \nu}$, where $T_\text{can}^{\mu \nu}$ is the canonical stress-energy tensor; however, as in the case of a CFT at zero temperature \cite{DiFrancesco:1997nk}, it can be shown that the broken Ward identities are not altered by replacing the canonical stress-energy tensor with the traceless, symmetric one: the only difference that can appear are divergences that will be removed in the normal ordering procedure.}, which is a well-defined operator in any local CFT, even if not Lagrangian. As we will see in Section \ref{sec: Superconformal BWI} for the superconformal case, the charges are related to the stress-energy tensor $T^{\mu \nu}(x)$ and the supercharge $G_{\alpha}^{\mu I}(x)$. The explicit realizations of the canonical currents associated with the symmetries of the Conformal Group are listed in Table \ref{tab:ExplicitThermalContributions}.

    By inserting the transformed action in the expression \eqref{eq: Correlator identity} and expanding at first order in the infinitesimal parameters $\omega_a$ we get 
    \begin{equation}\label{eq: Noether1}
    \begin{split}
        \sum_i  \left \langle \mathcal O_1(x_i) \ldots \delta \mathcal O_i(x_i) \ldots \mathcal O_n(x_n) \right \rangle_{\mathcal M} &=  \left \langle \delta S \  \mathcal O_1(x_1) \ldots \mathcal O_n(x_n) \right \rangle_{\mathcal M} \\ & = \int_{\cm} d^d x  \sqrt{g} \ \nabla_\mu \omega_a (x) \left \langle J_a^\mu (x) \mathcal O_1(x_1) \ldots \mathcal O_n(x_n) \right  \rangle_{\mathcal M} \ . 
    \end{split}
    \end{equation}
    Recalling that $\delta \mathcal O  (x) = -i \omega_a(x)\op G_a \mathcal O(x)$ and taking the functional derivatives in $\omega_{b}(y)$ both on the left and on the right side of the equation \eqref{eq: Noether1}, the expression reads 
    \begin{multline}
        -i \sum_{i}\delta(x_i-y)\left \langle \mathcal O_1(x_1) \ldots \op G_a\mathcal O_i(x_i) \ldots \mathcal O_n(x_n) \right \rangle_{\mathcal M} = \\ = \int _{\mathcal M}d^d x  \sqrt{g} \ \partial_\mu \delta(x-y) \ \left \langle J_{a}^{\mu}(x) \mathcal O_1(x_1) \ldots \mathcal O_n(x_n)\right \rangle_{\mathcal M} \ .
    \end{multline}
    The derivative in the $x$ coordinates can be replaced with a derivative in the $y$ coordinates up to a sign. We, therefore, get the un-integrated broken Ward identities 
    \begin{equation}\label{eq:UnintegratedBWI}
        i \sum_i \delta(x_i-y) \left \langle \mathcal O_1(x_1) \ldots \op G_a\mathcal O_i(x_i) \ldots \mathcal O_n(x_n) \right \rangle_{\mathcal M} = \nabla_\mu ^y\left \langle J_a^\mu(y) \mathcal O_1(x_1) \ldots \mathcal O_n(x_n) \right \rangle_{\mathcal M}  \ .
    \end{equation}
    It is crucial to notice that the current on the right-hand side of the equation \eqref{eq:UnintegratedBWI} iteratively acts on all the operators in the correlation function. For example, if we consider, on the thermal manifold $\mathcal M_\beta = S^1\times \mathbb R^{d-1}$, the dilatation broken Ward identity for a one-point function the equation \eqref{eq:UnintegratedBWI} specializes to
    \begin{equation}
        i \delta(x-y) \left \langle [\op D, \mathcal O](x) \right\rangle_\beta =  \frac{\partial}{\partial y^\mu}\left [ (y-x)_\nu \left \langle T^{\mu\nu}(y) \mathcal O(x) \right \rangle_\beta \right] \ ,
    \end{equation}
    where here and in the following we use $\langle \ \cdot\  \rangle_\beta = \langle\  \cdot\  \rangle_{\mathcal M_\beta}$ to indicate thermal correlators.
    This can be easily understood as the fact that we are integrating on the boundary of an infinite-dimensional ball, having as a center the position of the operator $\mathcal O$. On the other hand, this intuition perfectly agrees with the definition of the operator as the integration of the 0th component of the conserved current \cite{Zamolodchikov:1989hfa,Belavin:1984vu,Mussardo:2020rxh}; e.g. for dilatations
    \begin{equation} \label{eq: dil example}
        \op D  \, \mathcal O(x) = \int d^{d-1}y \ (x-y)_\nu T^{0\nu}(y)\mathcal O(x) \ .
    \end{equation}
    The un-integrated broken Ward identities, even if very explicit, can be simplified if we specify the manifold $\mathcal M$; as anticipated we are mainly interested in the thermal case, therefore from now on we will focus on this specific choice. A slightly different situation, namely $\mathcal M = T^2 \times \mathbb R^{d-2}$, is discussed in Section \ref{appendix:TorusCase}. In the thermal case $\mathcal M_\beta = S_1^\beta \times \mathbb R^{d-1}$, after integrating both the right and the left-hand sides of \eqref{eq: Noether1}, we get
    \begin{equation}\label{eq: general manifold Ward identity}
            i \sum_{i}\left \langle \mathcal O_1(x_1) \ldots \op G_a \mathcal O_i(x_i) \ldots  \mathcal O_n(x_n) \right \rangle_{\beta}  = \int d^{d-1} y\int_0^\beta d y_0  \ \frac{\partial}{\partial y^\mu }\left \langle J_{a}^\mu(y) \mathcal O_1(x_1) \ldots \mathcal O_n(x_n)\right \rangle_{\beta}\ .
    \end{equation}
    By replacing in the equation \eqref{eq: general manifold Ward identity} the thermal manifold $\mathcal M_\beta$ with the flat space $\mathbb{R}^{d}$, turning the volume integral into a boundary integral sets the right-hand side to zero (with the correct boundary conditions on the fundamental fields) and we recover the well known (un-broken) Ward identities 
    \begin{equation}
           i \sum_{i}\left \langle \mathcal O_1(x_1) \ldots \op G_a \mathcal O_i(x_i) \ldots  \mathcal O_n(x_n) \right \rangle_{\mathbb{R}^{d}}  =0 \ . \label{eq: flat WI}
    \end{equation}
    However, on a generic manifold $\cm$ the integration over the boundary $\partial \cm$ does not return a trivial result, hence the right-hand side of the equation \eqref{eq: general manifold Ward identity} can be interpreted as  a \emph{breaking term} correcting the Ward identity. In order to allow for explicit computations, we will parameterize the thermal circle $S^{1}_{\beta}$ with a coordinate $\tau \in \left[ 0, \beta \right)$, i.e. we impose  \footnote{In the following, Greek indices $\mu, \nu, \rho$ will run over all the directions of the thermal manifold (from 0 to $d-1$), while Latin indices $i,j,k$ will run over the spatial directions of the component $\mathbb{R}^{d-1}$ (from 1 to $d-1$). We work in Euclidean signature, so there is no difference between raised and lowered indices.} \begin{equation}
        \tau \equiv \tau+ \beta \ .
    \end{equation}
    We have to further impose boundary conditions for the fields: periodic and anti-periodic boundary conditions are assigned to bosons and fermions respectively.
	Now we can evaluate the breaking term on the thermal manifold.
 We can rewrite the right-hand side of the equation \eqref{eq: general manifold Ward identity} as the sum of two contributions: the first one is given by 
 \begin{equation}\label{eq: boundary term 1}
	    \int_{\mathbb R^{d-1}}d^{d-1}y \ \Braket{ \left[J_a^{0}(\beta,\vec y)-J_a^{0}(0,\vec y)\right]\mathcal O_1(x_1)\ldots \mathcal O_n(x_n)}_\beta \ ,
	\end{equation}
 and the second one by the integral over the boundary parallel to $S^{1}_{\beta}$
 \begin{equation}\label{eq: boundary term 2}
      \lim_{R\to \infty}\int_{0}^{\beta}d \tau \int_{S_1^{d-2}} d\Omega \ R^{d-2}n_i \left \langle J_a^{i}(\tau,R,\Omega)\mathcal O_1(x_1)\ldots \mathcal O_n(x_n) \right\rangle_\beta\ ,
 \end{equation}
 where $n_i$ is the unit vector fixing the orientation of the boundary and $d\Omega$ is the measure on the $(d-2)$-sphere of unitary radius. The term that appears in equation \eqref{eq: boundary term 2} can be discussed by invoking the clustering property 
 \begin{equation}\label{eq: clastering 1}
     \lim_{R \to \infty } \left \langle J_a^{i}(\tau, R,\Omega)\mathcal O(x_1) \ldots \mathcal O_n(x_n) \right \rangle_\beta =   \lim_{R\to \infty}\left \langle J_a^{i}(\tau,R,\Omega ) \right \rangle_\beta\left  \langle \mathcal O(x_1) \ldots \mathcal O_n(x_n)\right \rangle_\beta\ .
 \end{equation}
 To study the clustered correlator we can use translations invariance on the thermal manifold: no point in the spatial component $\mathbb{R}^{d-1}$ nor on the thermal circle $S^{1}_{\beta}$ should be regarded as preferred. Since the invariance under space-time translations is unbroken, every one-point function must be a constant. Consequently the term \eqref{eq: clastering 1} is constant in $R$, $\tau$, and $\Omega$ and therefore its integral over the thermal manifold is divergent unless the one-point function of the current is exactly zero. Therefore equation \eqref{eq: clastering 1} encodes a possible source of infrared divergence in the correlation function. The divergent term, if present, will be proportional to 
 \begin{equation}
      \langle \mathcal O(x_1) \ldots \mathcal O_n(x_n)\rangle_\beta \times \text{Vol}\left[\mathcal{M}_{\beta}\right] \ . \label{eq: bdy IR div}
 \end{equation}
 This suggests that the breaking term appearing in the broken Ward identities should be renormalized by subtracting all the possible infrared divergences: the immediate consequence is that the boundary term \eqref{eq: boundary term 2} is either zero or infrared divergent and therefore it does not contribute to the breaking term. This procedure is consistent (see Appendix \ref{appendix A}) and makes explicit some comments already present in literature \cite{Papadodimas2010}.
Let us now take into account the other boundary term in equation \eqref{eq: boundary term 1}. If we define the object
	\begin{equation}\label{eq: breaking term definition}
	 \Gamma^{\beta}_{a}(\vec{y})= J^{0}_{a}(\beta, \vec{y})- J^{0}_{a}(0, \vec{y}) \ , 
	\end{equation}
	we can rewrite the equation \eqref{eq: boundary term 1} as 
	\begin{equation} \label{eq: breaking term of the Ward identity}
		\int_{\mathbb{R}^{d-1}}d^{d-1}y \Braket{\Gamma^{\beta}_{a}(\vec{y})  \co_1(x_1) \dots \co_n(x_n)}_{\beta} \ .
	\end{equation}
    From this definition it is already clear that translations and spatial rotations are unbroken because of the periodicity of the current $J^\mu(\beta,\vec x) = J^{\mu}(0,\vec x)$. However, dilatations, rotations involving the time direction, and special conformal transformations can be broken since their respective currents are not periodic. Supersymmetry breaking is instead related to the anti-periodicity of the fermionic operators (which makes the supercurrent antiperiodic). \newline The final form of the broken Ward identities is then given by \begin{equation}\label{eq:GeneralBWI}
         i \sum_{i}\left \langle \mathcal O_1(x_1)\ldots \op  G_a \mathcal O_i(x_i) \ldots \ \mathcal O_n(x_n) \right \rangle_{\beta} = \int d^{d-1}y \ \left \langle \Gamma_a^{\beta}(\vec y) \mathcal O_1(x_1) \ldots \mathcal O_n(x_n) \right \rangle_\beta \ . 
    \end{equation}
    Further observe that among all the possible symmetry, the dilatation broken Ward identities gives information about the non-criticality of the theory and can be interpreted as a non-perturbative version of the Callan-Symanzik equation. Indeed it is easy to see that the dilatation broken Ward identities at finite temperature can be corrected by taking into account also the other scale present in the theory, namely the temperature. Indeed all correlators have to satisfy the differential equation \begin{equation}\label{eq:KallaS}
     \left (\op D + \beta \frac{\partial}{\partial \beta }\right)\left \langle \mathcal O_1(x_1) \ldots \mathcal O_n(x_n) \right \rangle_\beta = 0 \ .
 \end{equation}
    The derivation of the equation above is provided in Appendix \ref{Appendix1} and can be thought of as a simple dimensional analysis statement.
    \begin{table}[h!]
		\centering
		\renewcommand{\arraystretch}{1.5}
		\begin{tabular}{|c|c|}
			\hline
			Symmetry & Infinitesimal transformations: $i\omega_{a} \op G_{a}\cO{i}$ \\
			\hline
			\hline
			 Translations & $a^{\mu}\frac{\partial}{\partial x^{\mu}_i} \cO{i}$  \\
			\hline
			Dilatations & $b\left( x^{\mu}_i\frac{\partial}{\partial x^{\mu}_i} + \Delta_i\right)  \cO{i}$  \\
			\hline
			Rotations & $ c^{\mu \nu}\left(  -  x_{i,\mu}\frac{\partial}{\partial x^{\nu}_i}+x_{i,\nu}\frac{\partial}{\partial x^{\mu}_i}+i \op S_{\mu \nu} \right) \cO{i} $ \\
			\hline
			Special conf. tr. & $d^{\mu}\left( - x_i^2 \frac{\partial}{\partial x^{\mu}_i}+2  x_{i,\mu}^{\phantom{\nu}} x_{i}^{\nu}\frac{\partial}{\partial x^{\nu}_i}+ 2  x_{i,\mu} \Delta_i- 2i x^{\nu}_i \op S_{\mu \nu}^{\phantom{\nu}} \right)  \cO{i} $  \\
			\hline
		\end{tabular}
		\caption{\emph{Symmetries of the Global Conformal Group and their actions on the operator $\cO{i}$ living in an irreducible representation of the Lorentz Group \emph{\cite{DiFrancesco:1997nk}}. $a^{\mu}$ and $d^{\mu}$ are infinitesimal vectors, $b$ is an infinitesimal scalar and $c^{\mu \nu}$ an antisymmetric tensor; $\op S_{\mu \nu}$ represents the spin operator.}}
		\label{tab: currents and generators}
	\end{table}
    \newline Explicit checks of equation \eqref{eq:GeneralBWI} with the breaking terms of Table \ref{tab:ExplicitThermalContributions} were performed in two-dimensional CFTs and free theory in four space-time dimensions. The most relevant (and non-trivial) checks are provided in Appendix \ref{appendix A}. 
    \section{Conformal Group at finite temperature}\label{sec: Conformal group}
    By plugging in the general broken Ward identity \eqref{eq:GeneralBWI} the specific breaking terms associated with the symmetries of the Conformal Group (whose associated operators are listed in Table \ref{tab:ExplicitThermalContributions}), it is possible to produce specific constraints on the thermal correlation functions of a generic CFT at finite temperature. 
    In Table \ref{tab:ExplicitThermalContributions} we list for every symmetry in the Conformal Group the current and the associated breaking term $\Gamma^{\beta}_{a}(x)$, which can be computed by plugging the explicit expressions for the currents $J^{\mu}_{a}(x)$ in equation \eqref{eq: breaking term definition}.
  In the following, for the sake of notation clarity, we will set
	\begin{equation}
		\mathcal{O}_{1 \cdots n}(\bm{x}) = \co_1(x_1) \cdots \co_n(x_n) \ .
	\end{equation}
 \begin{table}[h!]
    \centering
	\renewcommand{\arraystretch}{1.5}
	\begin{tabular}{|c|c|c|}
		\hline
		Symmetry & Currents: $\omega_a J_a^{\mu}$ &Breaking terms: $\Gamma^{\beta}_{a}(0,\vec x)$ \\
		\hline
		\hline
		Translations  &$a_\nu T^{\mu \nu}$ & $0$ \\
		\hline
		Dilatations   & $b x_\nu T^{\mu \nu }$ & $\beta T^{00}(0,\vec x)$ \\
		\hline
		Spatial rot.  & $-c_{\, i }^{j}x_j T^{\mu i}+c_{i}^{\, j}x_j T^{\mu i}$ & $0$ \\
		\hline
		  Boosts &   $-c_{\, i}^{0}\tau T^{\mu i}+c_{i}^{\, 0}\tau T^{\mu i}$ & $\beta T^{0i}(0,\vec x)$ \\
        \hline
        S.c.t. on $S^{1}_{\beta}$ &
		$d \left(-x^2 \delta_{0 \nu}+2 \tau x_{\nu} \right)T^{\mu \nu} $ & 	$\beta \left[ (\beta-2 \tau) \, T^{00}(0,\vec x) +2  (x_i-y_i) T^{0i}(0,\vec x)\right]$ \\
        \hline 
        S.c.t. on $\mathbb{R}^{d-1}$ & $d^{i}\left(-x^2 \delta_{i \nu}+2 x_{i}x_{\nu} \right)T^{\mu \nu}$ & $\beta \left[ (\beta+2 \tau) \, T^{0j}(0,\vec x) +2  (x_i-y_i) T^{ij}(0,\vec x)\right]$\\ 
		\hline
	\end{tabular}
	\caption{\emph{Explicit $\Gamma_{a}^{\beta}(x)$ terms for the symmetries of the global Conformal Group. The $(\tau ,\vec y)$ coordinates locates the operator to which the breaking term is applied. Notice that the thermal contributions explicitly break the manifest $SO(d)$ invariance, but preserve the manifest $SO(d-1)$ invariance.}}
	\label{tab:ExplicitThermalContributions}
	\end{table}
	\paragraph{Translations} 
	The Ward identity associated with translations in every direction is left unbroken by the introduction of a finite temperature. The constraints are simply
	\begin{equation}\label{eq: unbroken translations}
		\sum_{r=1}^{n} \frac{\partial}{\partial \tau_r} \Braket{\mathcal{O}_{1 \cdots n}(\bm{x})}_{\beta}=0 \ , \qquad 	\sum_{r=1}^{n} \frac{\partial}{\partial x_r^i} \Braket{\mathcal{O}_{1 \cdots n}(\bm{x})}_{\beta}=0 \ .
	\end{equation}
	\paragraph{Dilatations}
	The Ward identity associated with dilatations is broken by the introduction of a finite temperature. In fact, the theory ceases to be scale-less: the effect is captured by the broken Ward identity
	\begin{equation}\label{eq: dilatation BWI}
		\sum_{r=1}^{n} \left(\tau_r \frac{\partial}{\partial \tau_r}+ x_r^{i} \frac{\partial}{\partial x_r^i}+ \Delta_r\right)  \Braket{\mathcal{O}_{1 \cdots n}(\bm{x})}_{\beta}= \beta \int d^{d-1}y \,\left \langle T^{00}(0,\vec y) \mathcal{O}_{1 \cdots n}(\bm{x})\right \rangle_\beta \ ,
	\end{equation}
    where $\Delta_{r}$ indicates the classical conformal dimension of the $r$-th operator.
	\paragraph{Rotations}
	From the list of breaking terms in Table \ref{tab:ExplicitThermalContributions}, we can notice that only boosts are associated to a broken Ward identity. Spatial rotations are associated to 
	\begin{equation} \label{eq: unbr rot wi}
		\sum_{r=1}^{n} \left(x_{r,i} \frac{\partial}{\partial x_r^j}-x_{r,j} \frac{\partial}{\partial x_r^i}-i \op S^{r}_{ij} \right)  \Braket{\mathcal{O}_{1 \cdots n}(\bm{x})}_{\beta}=0 \ ,
	\end{equation}
    where $\op S_{\mu \nu}^{r}$ denotes the application of the spin operator to the $r$-th operator. Boosts are instead associated with the following broken Ward identity 
	\begin{equation}\label{eq: rotation BWI}
		\sum_{r=1}^{n} \left(\tau_{r} \frac{\partial}{\partial x_r^i}-x_{r,i} \frac{\partial}{\partial \tau_r}+i \op S_{i0}^{r}\right)  \Braket{\mathcal{O}_{1 \cdots n}(\bm{x})}_{\beta}=\beta  \int d^{d-1}y \,\Braket{T^{0i}(0,\vec y) \mathcal{O}_{1 \cdots n}(\bm{x})}_\beta \ .
	\end{equation} 
	\paragraph{Special Conformal Transformations}
    We conclude the list with the broken Ward identities associated with the special conformal transformations. 
	  The broken Ward identity associated with special conformal transformations acting in the $S^{1}_{\beta}$ direction is 
	\begin{multline} \label{eq: sct1 gen}
		\sum_{r=1}^{n} \left[  \left(\tau_{r}^2- x^{i}_r x^{i}_r\right)  \frac{\partial}{\partial \tau_r}+2 \tau_r x^{i}_r \frac{\partial}{\partial x_r^i}+2 \tau_{r} \Delta_{r}+ 2 i x_{r}^{i} \op S^{r}_{i0} \right]  \Braket{\mathcal{O}_{1 \cdots n}(\bm{x})}_{\beta}=\\= \sum_{r= 1}^n \beta \left(\beta-2\tau_i\right)\int d^{d-1}y \left (\Braket{T^{00}(0,\vec y) \mathcal{O}_{1 \cdots n}(\bm{x})}_{\beta}+(x_i-y_i)\Braket{T^{0i}(0,\vec y) \mathcal{O}_{1 \cdots n}(\bm{x})}_{\beta}\right)\ ,
	\end{multline}
	while the broken Ward identity associated with the special conformal transformations acting in the $\mathbb{R}^{d-1}$ spacetime component is 
	\begin{multline}\label{eq: sct2 gen}
		\sum_{r=1}^{n} \left[  \left( \tau_{r}^2+x^{j}_r x^{j}_r\right)  \frac{\partial}{\partial x^i_r}-2 x_{r,i}x^{\mu}_{r} \frac{\partial}{\partial x^{\mu}_{r}}-2 x_{r,i} \Delta_{r}+2 i \tau_{r} \op S_{i0}^{r}+ 2i x_{r}^{j} \op S_{ij}^{r} \right]  \Braket{\mathcal{O}_{1 \cdots n}(\bm{x})}_{\beta}=\\= \sum_{r= 1}^n \beta \left(\beta+2\tau_i\right)\int d^{d-1}y \left (\Braket{T^{00}(0,\vec y) \mathcal{O}_{1 \cdots n}(\bm{x})}_{\beta}+(x_i-y_i)\Braket{T^{0i}(0,\vec y) \mathcal{O}_{1 \cdots n}(\bm{x})}_{\beta}\right) \ .
	\end{multline}
    \subsection{Hamiltonian, momentum and  (generalized) modularity} \label{ssec: ham and mom}
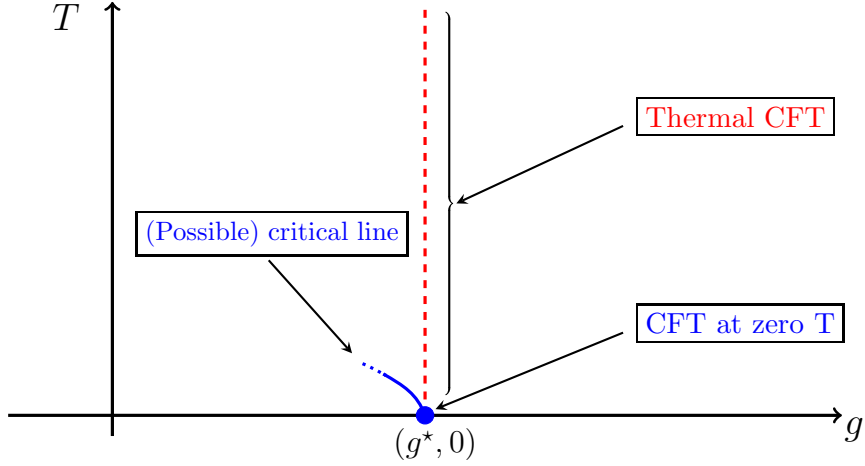
\begin{figure}[htb]
\centering
\begin{tikzpicture}[x=0.6pt,y=0.6pt,yscale=-1.30,xscale=1.30]
\draw[very thick, ->] (50,10)--(50,-200);
\draw[very thick, ->] (0,0)--(400,0);
\draw (20,-200) node [anchor=north west][inner sep=0.75pt]  [font=\Large]  {$T$};
\draw (400,0) node [anchor=north west][inner sep=0.75pt]  [font=\Large]  {$g$};
\draw (226,5) node [anchor=north east][inner sep=0.75pt]  [font=\large]  {$\left(g^\star, 0\right)$};
\draw (300,-55) node [anchor=north west][inner sep=0.75pt]  [font=\normalsize]  {\boxed{$\text{\textcolor{blue}{CFT at zero T}}$}};
\draw[thick, -stealth] (295, -40)--(205,-3);
%%%% CRITICAL LINE DRAWING %%%%%%%
\draw (60,-100) node [anchor=north west][inner sep=0.75pt]  [font=\small]  {\boxed{$\text{\textcolor{blue}{(Possible) critical line}}$}};
\draw[thick, -stealth] (125, -75)--(165,-30);
\draw (300,-155) node [anchor=north west][inner sep=0.75pt]
%%%%%%%%%%%%%%%%%%%%%%%%%%%%%%%%%%%
[font=\normalsize] {\boxed{$\text{\textcolor{red}{Thermal CFT}}$}};
\draw [thick, decorate, decoration = {calligraphic brace,mirror}] (210,-10) --  (210,-195);
\draw[thick, -stealth] (295, -140)--(215,-102.5);
\draw[very thick, dashed,red] (200,0)--(200,-200);
\filldraw[blue] (200,0) circle (2.5pt) node[anchor=north]{};
\draw [very thick,  blue] (200, 0) to[out=-110,in=30] (180,-20);
\draw [very thick, dotted, blue] (180, -20) to[out=30,in=20] (170,-25);
\end{tikzpicture}
\caption{\emph{A schematic representation of the $(g,T)$ plane, where $T$ is the temperature and $g$ is a coupling constant of the theory. The zero temperature CFT sits on the critical point $(g^\star,0)$, represented by a blue dot. In this paper, we study the thermal CFT, which can be represented as the QFT lying on the vertical dashed red line above the critical point. In this Figure, we inserted for clarity a pictorial representation of a possible blue line of critical points in the neighborhood of the zero-temperature fixed point, to distinguish the theories lying on it from the thermal CFT.}}
\label{fig:Scheme}
\end{figure}

    In this section we are going to derive the operatorial expression of the Hamiltonian and the momentum at finite temperature.  The energy spectrum that can be obtained from the diagonalization of such Hamiltonian corresponds to the energy spectrum of the theories that live on the vertical axis above the critical point at zero temperature in Fig. \ref{fig:Scheme}. 
    \newline
    It can be observed that the newly found broken Ward identities \eqref{eq: dilatation BWI} and \eqref{eq: rotation BWI} assume a very interesting physical interpretation when we introduce the Hamiltonian and spatial momentum\footnote{The minus sign is because we are using the convention for Euclidean theories, where $T_\text{Lorenzian}^{00} = i^2 T_\text{E
Euclidean}^{00} = -T_\text{Euclidean}^{00}$. The same for the $i$ in front of the definition of the momentum, $T_\text{Lorenzian}^{0i} = i T_\text{Euclidean}^{0i}$.}
    \begin{equation}\label{eq: Hamiltonian and momentum}
        H = -\int_{\mathbb R^{d-1}} d^{d-1}x \ \left(T^{00}(\tau,\vec x)+\frac{d-1}{d} \frac{b_T}{\beta^{d}} \right)\ , \qquad  P^i = i \int_{\mathbb R^{d-1}} d^{d-1}x \ T^{0i}(\tau,\vec x) \ ,
    \end{equation}
    where \cite{Iliesiu:2018fao} \begin{equation}
        \left \langle T^{00} \right\rangle_\beta =- \frac{d-1}{d} \frac{b_T}{\beta^{d}}\ .
    \end{equation}
    The latter equation, regarding the one-point function of the stress-energy tensor, can be obtained directly from equation \eqref{eq: dilatation BWI} when specialized to the case of one-point functions.
    By using the translational invariance of one-point functions, equation \eqref{eq: dilatation BWI} reads \begin{equation}\label{eq: D= H}
        \op H =-\frac{1}{\beta} \op D- E_0\ , 
    \end{equation}
    where $\op H$ is the Hamiltonian operator of the theory at finite temperature. In general, this theory should be thought of as a $(d-1)$-dimensional, interacting QFT at finite temperature \cite{Iliesiu:2018fao,Benjamin:2023qsc}. This means that the energies of the $(d-1)$-dimensional, interacting theory can be written in terms of the conformal spectrum of the $d$-dimensional conformal field theory; in particular, after having chosen as a basis $\ket{i} = \mathcal O_i \ket 0$ (where $\mathcal O_i$ is a scaling operator), it is possible to write down the matrix elements of the Hamiltonian as \begin{equation}\label{eq: D= H expli}
        \langle i | \op H |j \rangle_\beta =- \frac{\Delta_i}{\beta} \delta_{ij}-E_0\ ,
    \end{equation}
    where $\left\lbrace \Delta_{i} \right\rbrace$ is the spectrum of the dilatation operator when applied to operators (inserted in the origin). Those eigenvalues are zero temperature data and correspond to the conformal dimensions of the operator content of the zero-temperature CFT. Equation \eqref{eq: D= H expli} shows that these zero temperature data correspond to the energy spectrum at finite temperature, i.e. on the vertical red dashed axis in Fig. \ref{fig:Scheme}. For the sake of the heedful reader a clarification of what is depicted in Fig. \ref{fig:Scheme} is in order. A CFT at finite temperature is neither a critical point\footnote{It should be noticed that critical points at finite temperature exist: they are extremely interesting both theoretically and experimentally \cite{PhysRevLett.77.940,Chai:2020zgq,Chai:2020onq}, but they won't be discussed further in this project.} nor a fixed point of the RG flow. This is clear since dilatation invariance is broken. In fact, in the most general case, all the couplings run with the temperature. Nevertheless, if a critical point exists at zero temperature, it is possible to fix the couplings of the theory to be the critical ones at zero temperature and then consider the effects of the temperature on this theory. In this line of thinking the temperature plays the role of a scale and the observables (such as the \textit{thermal mass}) will change according to it. We show that the operatorial expressions of Hamiltonian and momentum in these non-trivial quantum theories, parameterized by the temperature, are consequences of the broken Ward identities. From the expression above one can read the energy spectrum of the theory at finite temperature\footnote{\label{footnote Lorentzian} The minus sign is due to the Euclidean signature. Switching to the Lorentzian one, the first contribution turns positive.}  \begin{equation}\label{eq: FT energy}
        E_n = -\frac{\Delta_n}{\beta} - E_0 \ ,
    \end{equation}
    where the $\Delta_n$ are ordered from the lowest to the highest one. The unitarity of the original CFT implies that the minimum value of $\Delta_n$ is given by $0$: this corresponds to the identity operator. $E_0$ is the energy of the thermal vacuum, which takes into account the Casimir energy of the system\footnote{The Casimir energy of a thermal system can be understood as a contribution given by the energy of the thermal gas: in fact, while at zero temperature $\Braket{ T^{00}} = 0$, at finite temperature $\Braket{ T^{00}}_{\beta} = - \left(1-\frac{1}{d}\right) \frac{b_T}{\beta^d}$, which returns the energy density $\mathcal E^{\text{gas}}_0 =  - \frac{d-1}{d} \frac{b_T}{\beta^{d-1}}$ \cite{Iliesiu:2018fao,PhysRevE.53.4414,Luo:2022tqy}.}. Equation \eqref{eq: FT energy} is perfectly in agreement with a trivial dimensional analysis: since the only scale in the theory is the temperature, i.e. $\beta$, then $E \propto 1/\beta$\footnote{$E_0$ is an exception since $E_0= \mathcal{E}^{\text{gas}}_0 \times \text{Vol}\left[\mathcal{M}_{\beta}\right]$.}. The non-triviality of equation \eqref{eq: FT energy} comes from the explicit coefficient in front of the temperature.
   Furthermore, in Appendix \ref{Appendix1} we show that the dilatation broken Ward identities at zero temperature can be corrected with a term proportional to $\beta \partial_\beta$, as already appeared in \cite{El-Showk:2011yvt}, \begin{equation} \label{eq: elshowk}
      \left ( \op D+ \beta \frac{\partial}{\partial \beta}\right)\langle \mathcal O_1(x_1) \ldots \mathcal O_n(x_n) \rangle_\beta = 0 \ ,
   \end{equation}
   where $\partial_\beta$ plays the role of the Hamiltonian, and it is directly related to $-\op D/\beta$ via the relation above confirming the above identification of the Hamiltonian with the (properly rescaled) dilatation operator. We show explicitly in Appendix \ref{Appendix1} that the dialtation broken Ward identity is equivalent with equation \eqref{eq: elshowk}.
    
    Similarly to the dilatation operator, the boost operator $\op L_{i}$, is related to momentum operators via equation \eqref{eq: rotation BWI}. Explicitly  
    \begin{equation}\label{eq: S= P}
       \op P_i = \frac{i}{\beta}  \op L_{i}   \ , 
    \end{equation}
    where $\op P_i$ is the spatial momentum operator of the theory at finite temperature. The physical interpretation of the  equations \eqref{eq: D= H} and \eqref{eq: S= P} is that the dilatation operator generates, at finite temperature, the time translations (where here the time direction is the compactified direction on $S^1_{\beta}$), whereas the boost operator generates the spatial translations. 
    \newline As already observed in \cite{Iliesiu:2018fao}, the definitions of Hamiltonian and momentum in \eqref{eq: Hamiltonian and momentum} are the most natural ones, interpreting the system as a finite temperature system. However, from the Kaluza-Klein perspective, this is not the most natural way of defining Hamiltonian and momentum; in the latter case, it is more natural to compactify a space direction and not the time direction. Defining $x^1\in \mathbb R$ as the time direction
    \begin{equation}
         H = -\int_0^\beta d \tau \int_{\mathbb R^{d-2}} d^{d-2}x \ \left(T^{11}(\tau,\vec x)+\frac{1}{d} \frac{b_T}{\beta^{d}} \right)\ , \quad  P^i = i \int_0^\beta d\tau \int_{\mathbb R^{d-2}} d^{d-2}x \ T^{1i}(\tau,\vec x) \ ,
    \end{equation}
    where $i = 0,2,3,4,\ldots$ 
    \newline 
    \newline There is also a third possibility we are not exploring in this paper concerning the CFT at finite volume (the Conformal Group is broken also in this case). In the language of this paper, it can be defined as a conformal field theory on the manifold $\mathbb R \times S_{R}^{d-1}$, where $\mathbb R$ represents the time direction and the space directions are compactified on a $(d-1)$-sphere of radius $R$.
    In two dimensions,  Hamiltonian and momentum of finite volume theories are well known in the literature \cite{Cardy:1989da,Cardy:1991kr,Mussardo:2020rxh} and can be checked explicitly, by exploiting the conformal map between the plane $\mathbb R^2 \simeq \mathbb C$ and the cylinder $\mathbb R \times S^1_{R}$. Generalizations to higher dimensions are also provided in literature together with impressive applications \cite{Iliesiu:2018fao,Hogervorst:2013sma,Hogervorst:2014rta}.
    In this perspective, equations \eqref{eq: D= H}, \eqref{eq: S= P} are curiously and intriguingly similar (the only differences concern the Euclidean/Lorentzian conventions) to the Hamiltonian and momentum at finite volume, hinting at a possible relation between finite temperature and finite volume systems. This fact is understood in two dimensions, since it is a consequence of the modular invariance \cite{Cardy:1986ie,DiFrancesco:1997nk,Mussardo:2020rxh}, but  it is not completely clear in higher dimensions. As already proposed in literature \cite{Berg:2023pca,Downing:2023uuc}, we can speculate on the possibility to \textit{generalize the modular invariance} in higher dimensions. One way is to consider the conformal field theory on $S_1^\beta \times S_{d-1}^{R}$ in which we choose the time direction to be defined by $0 \le \tau \le \beta$ and the spatial direction to be $0 \le x^i \le R$. The latter geometry is related to $\mathbb R^{d}$ via conformal maps; the thermal CFT can be derived by sending $R \to \infty$ while the finite temperature version of the same theory can is describe by the limt $\beta \to \infty$. Therefore the conformally equivalent geometry $S_1^\beta \times S_{d-1}^{R}$ contains information both of the thermal and the finite volume theory.
    \newline Furthermore the consistency of the theory on the thermal manifold can be thought of as a \textit{generalized modular invariance} \cite{El-Showk:2011yvt} and this fact seems to be captured for the first time from this computation, since the Hamiltonian computed for the finite temperature system is the same Hamiltonian (up to constant factors) known for the finite volume system.

     The Hamiltonian of a finite temperature CFT can be written in terms of the dilatation operator $\op D$ of the zero-temperature CFT, and how the momentum operator of the finite temperature system can be expressed in terms of the boost operator of the zero-temperature CFT. This simple observation seems to suggest that some finite temperature quantities could be recovered by considering zero-temperature CFT data, through the, previously derived, broken Ward identities. 
  \newline One very important finite temperature quantity in which we are interested is the \emph{thermal mass} of the theory. This mass is not a physical mass, but it can be defined as the exponential factor that rules the behavior of the two-point functions at spatial infinity, i.e. in the limit $|\vec x|\to \infty $  
    \begin{equation}
        \left \langle \phi(\tau,\vec x) \phi(0)\right \rangle_\beta \simeq \left \langle \phi(0)\right \rangle_\beta^2+ O \left(e^{-m_{th}|\vec x|}\right) \ ,
    \end{equation}
    where $m_{th}$ is the thermal mass. When $\vec x = (x,0,0,\ldots)$, we have \begin{equation}
        \left \langle \phi(0) \phi(\tau,x) \right \rangle_\beta = \left \langle \phi(0) e^{i \op H \tau- \op P_1 x } \phi(0) \right \rangle_\beta \ ,
    \end{equation}
    where $\op H$ and $\op P_1$ are the operators defined in the previous Section. Observe that the definition of the thermal mass is simply the first non-zero eigenvalues of $\op P_1$, which is a well-defined differential operator. In particular, it is possible to choose the basis of the Hilbert space $\ket{\mathcal O_{\Delta,J}} =\mathcal O_{\Delta,J}(0) \ket{0}$ thanks to state-operator correspondence. Then the thermal mass is  given by 
    \begin{equation}
       m_\text{th} = \frac{1}{\beta} \min_{\Delta,J} \left \langle \mathcal O_{\Delta,J}|i \op L_1|\mathcal O_{\Delta,J} \right \rangle  = \frac{1}{\beta}  \min_{\Delta,J} \lim_{y \to \infty } y^{2\Delta} \left \langle \mathcal O_{\Delta,J}(y) i \op S_{01}\mathcal O_{\Delta,J}(0)\right \rangle \ ,
    \end{equation}
    where it is understood that the minimum excludes zero eigenvalues and the correlators are zero temperature correlators.
    \subsubsection{An implicit version of the Cardy formula}
    In Appendix \ref{Appendix1} we presented a different derivation of broken Ward identities, leading to equation \eqref{eq:KallaS}, and we observed that when applied to one point functions, dilatations Ward identities imply \begin{equation}
      \frac{\partial}{\partial \beta }\left \langle \mathcal O(0)\right \rangle_\beta = -\frac{1}{\beta}\int d^{d-1}x \left \langle T^{00}(0,\vec x) \mathcal O(0)\right \rangle_\beta \ .
    \end{equation}
    This formula already appeared in literature \cite{El-Showk:2011yvt} and it has a very concrete simple interpretation. The temperature is related to the radius of the thermal circle, or equivalently to the value of the metric component $g_{00}$. Since we can think at $g_{00}$ as the \text{source} of $T^{00}$, then a small change in the temperature is equivalent to the insertion of the zero-zero component of the stress-energy tensor. This physical interpretation is extremely useful, in particular considering the specific case in which \begin{equation}\label{eq: Cardy formula}
      \frac{\partial}{\partial \beta }\langle T^{00}(0)\rangle_\beta = -\frac{1}{\beta}\int d^{d-1}x
      \left \langle T^{00}(x) T^{00}(0) \right \rangle_\beta \ .
    \end{equation}
  It is natural to think \eqref{eq: Cardy formula} as the differential equation generalizing the Cardy Formula in every spacetime dimension \cite{El-Showk:2011yvt}. Indeed using thermodynamic relations it is possible to compute the entropy of the system as a function of the $b_T$ coefficient, which plays the same role of the central charge $c$ in two dimensions\footnote{Indeed in two dimensions $b_T \propto c$.}. The $b_T$ coefficient is implicit in equation \eqref{eq: Cardy formula} and it is not known in general, apart from exceptional cases, leaving the generalization of the Cardy formula in higher dimensions an interesting problem to be addressed in the future \cite{DiPietro:2014bca,Assel:2015nca,Mukhametzhanov:2019pzy,Carlip:2000nv,Brevik:2004sd,Wang:2001bf,Benjamin:2023qsc}. 
    \subsection{Thermal one-point functions} \label{ssec: Thermal 1-point Functions}
	We consider a generic thermal one-point function $\langle \mathcal O(x)\rangle_\beta$, where $\mathcal{O}(x)$ is a local operator. The unbroken Ward identities \eqref{eq: unbroken translations} are of fundamental importance: in our case, they specialize to
	\begin{equation}
		\frac{\partial}{\partial \tau} \left \langle \mathcal O(\tau,\vec x)\right \rangle_{\beta}=0 \ ,  \qquad \frac{\partial}{\partial x^i} \left \langle \mathcal O(\tau,\vec x)\right \rangle_{\beta}=0 \ .
	\end{equation}
	Thanks to these equations, we conclude that a thermal one-point function (of a local operator) is a constant all over the thermal manifold. Once we acknowledge this, the other Ward identities can be significantly simplified. First of all, the unbroken Ward identity describing spatial rotations simply becomes
    \begin{equation}\label{eq: 1pt functions unbroken rot}
     \left \langle \op S_{ij} \mathcal O(x) \right \rangle_\beta = 0 \ ;
    \end{equation}
    dilatations and boosts are encoded into simplified broken Ward identities
	\begin{align} \label{eq: 1pt functions dil}
			\Delta_{\co}\langle \mathcal O(x)\rangle_\beta  &=  \beta \int d^{d-1}y \, \Braket{ T^{00}(0,\vec y) \mathcal O(x)}_\beta \ , \\
    i\langle \op S_{i0}  \mathcal O(x)\rangle_\beta &=  \beta \int d^{d-1}y \, \Braket{T^{0i}(0,\vec y)\mathcal O(x)}_\beta \ . \label{eq: 1pt functions rot}
	\end{align}
    The broken Ward identity associated with special conformal transformations in the $S^{1}_{\beta}$ direction simplifies to
    \begin{multline}
		    2 \tau \Delta_{\mathcal{O}}\Braket{\mathcal{O}(\tau,\vec{x})}_{\beta}+2  i x^{i} \Braket{\op S_{i0}\mathcal{O}(\tau,\vec{x})}_{\beta}=\beta \left(\beta-2\tau\right) \int d^{d-1}y \Braket{T^{00}(0,\vec y)\mathcal{O}(\tau,\vec{x})}_{\beta}+\\+\beta \int d^{d-1}y \ (x_i-y_i)\Braket{T^{0i}(0,\vec y)\mathcal{O}(\tau, \vec{x})}_{\beta} \ ,
	\end{multline}
    while the one in the $\mathbb{R}^{d-1}$ component is 
    \begin{multline}
		-2 x_{i} \Delta_{\mathcal{O}}\Braket{\mathcal{O}(\tau, \vec{x})}_{\beta}+ 2 i \tau \Braket{\op S_{i0}\mathcal{O}(\tau, \vec{x})}_{\beta}=\beta \left(\beta+2\tau\right) \int d^{d-1}y \Braket{T^{0i}(0,\vec y)\mathcal{O}(\tau,\vec{x})}_{\beta}+\\+\beta \int d^{d-1}y \ (x_j-y_j)\Braket{T^{ij}(0,\vec y)\mathcal{O}(\tau, \vec{x})}_{\beta} \ .
	\end{multline}
 To summarize, we conclude that the unbroken and broken Ward identities of the Global Conformal Group constrain the thermal one-point functions to be kinematically constant. From the point of view of their tensorial structure, they must be $SO(d-1)$-invariant tensors. All the other features can be extracted in principle from the equations \eqref{eq: 1pt functions dil} and \eqref{eq: 1pt functions rot}. As anticipated, we will now focus on the thermal one-point functions of scalar, vector, and rank-two tensor local operators.
	\paragraph{Scalar operators}
	Scalar operators are particularly easy to study since $\Braket{\op S_{0i} \mathcal O(x) }_\beta = \Braket{ \op S_{ij}\mathcal O(x) }_\beta=0$ identically. The rotation broken Ward identity simply becomes 
	\begin{equation}
        \int d^{d-1}y \, \Braket{T^{0i}(0,\vec y)\mathcal O(x) }_\beta =0 \ .
	\end{equation}
	which is satisfied, since for $SO(d-1)$ invariance the integrand has to be odd in one spatial coordinate. The only meaningful constraint on the thermal one-point function of a scalar operator $\co(x)$ is given by the broken Ward identity \eqref{eq: 1pt functions dil}
	\begin{equation} \label{eq: 1pt thermal function}
		\Braket{ \co(x) }_{\beta}=\frac{\beta}{\Delta_{\co}} \int d^{d-1}y \Braket{T^{00}(0,\vec y)\co(x)}_{\beta} \ .
	\end{equation}
    As already noticed this equation is extremely interesting and can give, in principle, non-trivial constraints on structure constants of the zero-temperature theory.
	\paragraph{Vector operators} 
	When we consider a local vector operator $\mathcal O^\mu(x)$, the Ward identities associated with rotations are not trivial anymore. We need to evaluate explicitly the functions $\left \langle \op S_{0i} \mathcal O^{\mu}(x) \right \rangle_\beta$ and $\left \langle \op S_{ij}  \mathcal O^{\mu}(x)\right \rangle_\beta$. First of all, we specialize the generators $\op S_{ij}$ and $\op S_{0i}$ to the vector representation of  $SO(d)$
	\begin{equation}
		\tensor{\left(\op S_{ij} \right)}{^\mu_{\nu}}=i\left(\delta^{\mu}_{i}\delta^{\phantom{\mu}}_{j \nu}-\delta^{\mu}_{j}\delta^{\phantom{\mu}}_{i \nu}\right) \ , \qquad \tensor{\left(\op S_{0i} \right)}{^\mu_{\nu}}=i \left(\delta^{\mu}_{0}\delta^{\phantom{\mu}}_{i \nu}-\delta^{\mu}_{i}\delta^{\phantom{\mu}}_{0 \nu}\right) \ . 
	\end{equation}
	We can now explicitly write down the unbroken and broken Ward identities
	\begin{align}
		\left \langle \op S_{ij} \mathcal{O}^{\mu}(x)\right \rangle_\beta&=i \left(\delta^{\mu}_{i}\Braket{\co_{j}(x)}_{\beta}-\delta^{\mu}_{j}\Braket{\co_{i}(x)}_{\beta}\right) =0 \ ,  \label{eq: 1pt vec rot 1}\\ \left \langle \op S_{0i} \mathcal{O}^{\mu}(x)\right \rangle_\beta&=i \left(\delta^{\mu}_{0}\Braket{\co_{i}(x)}_{\beta}-\delta^{\mu}_{i}\Braket{\co_{0}(x)}_{\beta}\right)= -i \beta \int d^{d-1}y \Braket{T^{0i}(0,\vec y)\co^{\mu}(x)}_{\beta} \label{eq: 1pt vec rot 2} \ .
	\end{align}
	Let us start by solving the equation \eqref{eq: 1pt vec rot 1}. If we set $\mu=0$, then the identity is trivially satisfied; if we set $\mu=k$, then we can multiply both sides by $\delta^{i}_{k}$, obtaining
	\begin{equation} \label{eq: spatial components}
		\left( d-2\right) \Braket{\co_{j}(x)}_{\beta}=0 \ .
	\end{equation}
	This equation is always satisfied in $d=2$, hence it doesn't provide any additional constraint; in $d>2$, it returns 
	\begin{equation}
		\Braket{\co_{i}(x)}_{\beta}=0
	\end{equation}
	which is also a solution of the unbroken Ward identity \eqref{eq: 1pt vec rot 1}. This was to be expected for symmetry reasons. Moreover, it simplifies the equation \eqref{eq: 1pt vec rot 2}
	\begin{equation}
		\delta^{\mu}_{i}\Braket{\co^{0}(x)}_{\beta}= \beta \int d^{d-1}y \Braket{T^{0i}(0,\vec y)\co^{\mu}(x)}_{\beta}\ .
	\end{equation}
	By setting $\mu=0$, we get an equation satisfied for symmetry reasons; instead, by setting $\mu=j$ and multiplying by $\delta^{i}_{j}$ we get an explicit expression for the $0$-th component
	\begin{equation}
		 \Braket{\co^{0}(x)}_{\beta}= \frac{\beta}{d-1} \int d^{d-1}y \Braket{T^{0i}(0,\vec y)\co^{i}(x)}_{\beta} \ .
	\end{equation}
 The dilatations broken Ward identity is 
 \begin{equation}
		\Braket{\co^{\mu}(x)}_{\beta} = \frac{\beta}{\Delta_{\co}} \int d^{d-1}y \Braket{T^{00}(0,\vec y)\co^{\mu}(x)}_{\beta} \ .
	\end{equation}
	If we consider $\mu=i$ we get an identity for symmetry reasons, while if we consider $\mu=0$, we get a second constraint on the $0$-th component 
	\begin{equation}
		\Braket{\co^{0}(x)}_{\beta} = \frac{\beta}{\Delta_{\co}} \int d^{d-1}y \Braket{T^{00}(0,\vec y)\co^{0}(x)}_{\beta} \ .
	\end{equation}
	In conclusion, the most general statement we can produce only by applying the unbroken and broken Ward identities is
	\begin{align}
		\Braket{\co^{i}(x)}_{\beta}&=0 \ , \\ \Braket{\co^{0}(x)}_{\beta} &= \frac{\beta}{\Delta_{\co}} \int d^{d-1}y \Braket{T^{00}(0,\vec y)\co^{0}(x)}_{\beta}= \frac{\beta}{d-1} \int d^{d-1}y \Braket{T^{0i}(0,\vec y)\co^{i}(x)}_{\beta} \ .
	\end{align}
	In Section \ref{subsec: constraints in the OPE regime}, we will show $\left \langle \mathcal O^{0}(x)\right \rangle_\beta=0$ if $\mathcal{O}^{\mu}$ lives in the OPE of two scalar local operators.
	\paragraph{Rank-two tensor operators} \label{sec Two rank operators}
	We focus on a local rank-two tensor operator $\co^{\mu \nu}(x)$. The unbroken Ward identity associated with spatial rotations can be written as
	\begin{equation}
		\left \langle \op S_{ij}  \mathcal O_{\mu \rho}(x) \right \rangle_\beta =i\left(\delta_{\mu i}\Braket{\co_{j \rho}(x)}_{\beta}-\delta_{\mu j}\Braket{\co_{i \rho}(x)}_{\beta}+\delta_{i \rho}\Braket{\co_{\mu j}(x)}_{\beta}-\delta_{j\rho}\Braket{\co_{\mu i}(x)}_{\beta}\right)	=0 \ .
	\end{equation}
	If we set $\mu=\rho=0$, we get a trivial identity, so we simply set
 \begin{equation}
     \Braket{\co_{0 0}(x)}_{\beta}=a \ , 
 \end{equation}
 where $a$ is a constant. By setting $\mu=k$ and $\rho=0$, we get an equation similar to \eqref{eq: spatial components}
	\begin{equation}
		(d-2)\Braket{\co_{j 0}(x)}_{\beta}=0 \ , 
	\end{equation}
	which again is trivially solved if $d=2$, while in higher dimensions it produces the following constraint
	\begin{equation}
		\Braket{\co_{i 0}(x)}_{\beta}=0 \ .
	\end{equation}
	We can get $\Braket{\co_{0 j}(x)}_{\beta}=0 $ in a similar way. Finally, we want to constraint $\Braket{\co_{i j}(x)}_{\beta}$ by choosing $\mu=k$ and $\rho=l$
	\begin{equation}
		\langle \op S_{ij} \mathcal O_{kl}\left(x\right)\rangle_\beta =i\left(\delta_{k i}\Braket{\co_{j l}(x)}_{\beta}-\delta_{k j}\Braket{\co_{i l}(x)}_{\beta}+\delta_{i l}\Braket{\co_{k j}(x)}_{\beta}-\delta_{j l}\Braket{\co_{k i}(x)}_{\beta}\right)	=0 \ .
	\end{equation}
	This equation is trivially satisfied in $d=2$ since the only existing rotation is the one generated by $\op S_{01}$. In higher dimension, the  solution is given by the only $SO(d-1)$-invariant tensor with two indices
	\begin{equation}\label{eq: b definition }
		\Braket{\co_{i j}(x)}_{\beta}=b \delta_{ij} \ ,
	\end{equation}
	where $b$ is a constant.
	\newline We conclude that the thermal one-point function $\Braket{\co^{\mu \nu}(x)}_{\beta}$ is completely determined by the constants $a$ and $b$. We can use the other Ward identities to set constraints on $a$ and $b$. In particular, by using the dilatation broken Ward identity, the only non-trivial constraints are
	\begin{align}
		a&=\Braket{\co^{00}(x)}_{\beta}=\frac{\beta}{\Delta_{\co}}\int d^{d-1}y\Braket{T^{00}(0,\vec y) \co^{00}(x)}_{\beta}\ , \\ b&=\frac{\beta \delta_{ij}}{d-1}\Braket{\co^{ij}(x)}_{\beta}=\frac{\beta \delta_{ij}}{(d-1)\Delta_{\co}}\int d^{d-1}y\Braket{T^{00}(0,\vec y) \co^{ij}(x)}\ .
	\end{align}
    The broken Ward identity associated with boosts establishes a relationship between the two constants: the identity reads
	\begin{equation} \label{eq: rot bwi 2rank tensors}
	\delta_{\mu 0}\Braket{\co_{i \rho}(x)}_{\beta}-\delta_{\mu i}\Braket{\co_{0 \rho}(x)}_{\beta}+\delta_{0 \rho}\Braket{\co_{\mu i }(x)}_{\beta}-\delta_{i \rho}\Braket{\co_{\mu 0 }(x)}_{\beta}=-\int d^{d-1}y \Braket{T^{0i}(0,\vec y)\co_{\mu \rho}(x)}_{\beta}
	\end{equation}
	By choosing $\mu=0$ and $\rho=i$, we extract the only non-trivial identity
	\begin{equation} \label{eq: a and b constraint}
		b=a-\frac{1}{d-1} \int d^{d-1}y \Braket{T^{0i}(0,\vec y)\co^{0 i}(x)}_{\beta}.
	\end{equation}
    In Section \ref{subsec: constraints in the OPE regime}, we will show that, if $\mathcal{O}^{\mu \nu}$ lives in the OPE of two scalar operators, $a$ and $b$ must comply to an additional constraint that sets the thermal one-point function to be traceless. 
\newline 
    We conclude by remarking how all the results we obtained are in agreement with the general result of \cite{Iliesiu:2018fao} 
    \begin{equation}\label{eq: one point functions}
        \langle \mathcal O^{\mu_1\ldots \mu_J}(x)\rangle_{\beta} = \frac{b_{\mathcal O}}{\beta^{\Delta_{\mathcal O}}} \left(e^{\mu_1}\cdots e^{\mu_J}-\text{traces}\right) \ ,
    \end{equation}
    where $e^{\mu}$ is a unit vector in the compactified direction of $S^1_{\beta}$. The authors of \cite{Iliesiu:2018fao} got the result from a pure symmetry-driven argument; in our work symmetries' constraints are encoded in the broken and unbroken Ward identities and, even if these differential equations are in general complicated, in the OPE regime they return the same results of \cite{Iliesiu:2018fao}, as expected. If the broken Ward identities can constrain the correlators away from the OPE regime it remains an open problem we are not going to address in this paper.
	\subsection{Thermal scalar two-point functions} \label{ssec: Thermal Scalar 2-points Functions}
	The natural step following the analysis of the thermal one-point functions is to consider thermal two-point functions. Two-point functions, differently from one-point functions, retain an explicit kinematic dependence. The goal of this Section is to formulate a bootstrap problem for the two-point function of two \emph{scalar} local operators $\mathcal O_1(x_1)$ and $\mathcal O_2(x_2)$. Working with scalars, the set of broken and unbroken Ward identities significantly simplifies, motivating our choice. In order to employ a more compact notation, we set
	\begin{equation}
		\Braket{\cO{1}\cO{2}}_{\beta}=f_{\beta}\left(\tau_1,\tau_2, \vec{x}_1, \vec{x}_2 \right) \ .
	\end{equation}
 \subsubsection{Broken and unbroken Ward identities constraints} \label{ssec: bwi 2pts}
	The easiest constraint is given by the unbroken translations Ward identity
	\begin{align}
		\frac{\partial}{\partial \tau_1} f_{\beta}\left(\tau_1,\tau_2, \vec{x}_1, \vec{x}_2 \right)+\frac{\partial}{\partial \tau_2} f_{\beta}\left(\tau_1,\tau_2 ,\vec{x}_1, \vec{x}_2 \right)&=0 \ , \\	\frac{\partial}{\partial x_1^i} f_{\beta}\left(\tau_1,\tau_2, \vec{x}_1, \vec{x}_2 \right)+\frac{\partial}{\partial x_2^i} f_{\beta}\left(\tau_1,\tau_2, \vec{x}_1, \vec{x}_2 \right)&=0 \ .
	\end{align}
	These equations are solved in full generality by
	\begin{equation}
		f_{\beta}\left(\tau_1,\tau_2, \vec{x}_1, \vec{x}_2 \right)=f_{\beta}\left(\tau_1-\tau_2, \vec{x}_1 - \vec{x}_2 \right)=f_{\beta}\left(\tau_{12}, \vec{x}_{12} \right) \ .
	\end{equation}
	We can now consider the constraints coming from applying the broken and unbroken rotations Ward identity. Since we are focusing on scalar local operators, the equations simplify. The unbroken Ward identity \eqref{eq: unbr rot wi} reads
	\begin{equation} \label{eq: rot1 2pts funct}
		\left( x_{12,i}  \frac{\partial}{\partial x_{12}^j}-x_{12,j} \frac{\partial}{\partial x_{12}^i}\right)  f_{\beta}\left(\tau_{12}, \vec{x}_{12} \right)=0 
	\end{equation}
	and it is solved by 
	\begin{equation}
		f_{\beta}\left(\tau_{12}, \vec{x}_{12} \right)=f_{\beta}\left(\tau_{12}, \left| \vec{x}_{12}\right|  \right) \ ; \label{eq: coord dep}
	\end{equation}
	The broken Ward identity \eqref{eq: rotation BWI} instead reads
	\begin{equation} \label{eq: rot2 2pts funct}
		\left( \tau_{12}  \frac{\partial}{\partial x_{12}^i}-x_{12,i} \frac{\partial}{\partial \tau_{12}}\right)  f_{\beta}\left(\tau_{12},  \left| \vec{x}_{12}\right| \right)=\beta \int d^{d-1}y \Braket{T^{0i}(0,\vec y)\cO{1}\cO{2}}_{\beta} \ ,
	\end{equation}
	which represents a new non-trivial constraint on the two-point function. Interestingly, the right-hand side of equation \eqref{eq: rot2 2pts funct} depends only on the difference $\tau_{12}$. This observation allows us to highlight that the distance between two operators on the thermal circle can be measured in two different ways, corresponding to the two arcs identified by the two points $\tau_1$ and $\tau_2$. The independence of the correlation function on the choice of the arc is a consequence of the KMS condition. When introducing a third operator, such as the stress-energy tensor in equation \eqref{eq: rot2 2pts funct}, the aforementioned choice generates two possibilities: one where the stress-energy tensor is between the two operators, i.e., the distance is measured on the arc including the third operator, and another where it is not. However, the KMS condition ensures the equivalence of these two choices. The dilatations broken Ward identity provides an additional non-trivial constraint 
	\begin{equation} \label{eq: dil 2pts funct}
		\left(\tau_{12} \frac{\partial}{\partial \tau_{12}}+ x_{12}^{i} \frac{\partial}{\partial x_{12}^i}+\Delta_1+\Delta_2 \right) f_{\beta}\left(\tau_{12},  \left| \vec{x}_{12}\right| \right) = \beta \int d^{d-1}y \Braket{T^{00}(0,\vec y)\cO{1}\cO{2}}_{\beta} \ .
	\end{equation}
We complete the exposition of the complete set of constraints on the thermal scalar two-point functions coming from the Ward identities with the special conformal transformations. The equations \eqref{eq: sct1 gen} and \eqref{eq: sct2 gen} in this case specialize in complicated equations, however the breaking terms of the special conformal transformations can be written in terms of breaking terms of dilatations and rotations. Therefore, also considering   $\beta - \tau=\tau$, we do not expect any additional constraints coming from special conformal transformations.
    \subsubsection{Additional consistency constraints}\label{subsec: Other consistency conditions}
\begin{figure}[t]%[htb]
\centering
\begin{tikzpicture}
[x=0.6pt,y=0.6pt,yscale=-1,xscale=1]
\draw[->,thick] (-200,0)--(200,0);
%\draw[->,thick] (0,130)--(0,-130);
\draw (155,-70)--(155,-55);
\draw (155,-55)--(175,-55);
\draw[thick,decorate, decoration ={zigzag},red] (-130,0)--(-200,0);
\draw[thick,decorate, decoration ={zigzag},red] (130,0)--(200,0);
\draw[thick,decorate, decoration ={zigzag},red] (-50,0)--(50,0);
\filldraw[red] (90,0) circle (2pt) node[anchor=north]{};
\filldraw[red] (-130,0) circle (2pt) node[anchor=north]{};
\filldraw[red] (130,0) circle (2pt) node[anchor=north]{};
\filldraw[red] (50,0) circle (2pt) node[anchor=north]{};
\filldraw[red] (-50,0) circle (2pt) node[anchor=north]{};
\filldraw[red] (-90,0) circle (2pt) node[anchor=north]{};

\draw (-160,10) node [anchor=north west][inner sep=0.75pt]  [font=\large]  {$-1/r$};
\draw (110,10) node [anchor=north west][inner sep=0.75pt]  [font=\large]  {$1/r$};
\draw (-70,10) node [anchor=north west][inner sep=0.75pt]  [font=\large]  {$-r$};
\draw (45, 10) node [anchor=north west][inner sep=0.75pt]  [font=\large]  {$r$};
\draw (-110, -30) node [anchor=north west][inner sep=0.75pt]  [font=\large]  {$-\tilde \omega$};
\draw (85, -30) node [anchor=north west][inner sep=0.75pt]  [font=\large]  {$\tilde \omega$};
\draw (160,- 70) node [anchor=north west][inner sep=0.75pt]  [font=\large]  {$\omega$};

\end{tikzpicture}
\caption{Scheme of the analyticity condition of the two-point function $f(\tau,|\vec x|) = f(r/\omega,r \omega )$ in the $\omega$ plane. As shown in \cite{Iliesiu:2018fao}, the two-point function must be analytic everywhere outside the poles and the cuts drawn in the $\omega$ plane.}
\label{Analiticity}
\end{figure}
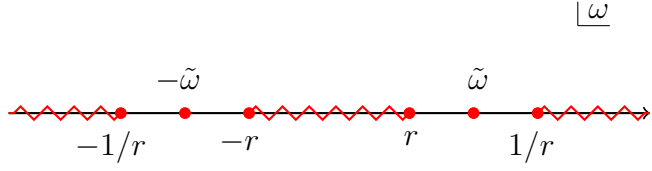

     In the previous Section, we discussed how thermal scalar two-point functions can be constrained by enforcing the broken and unbroken Ward identities. In this Section, we recap the other important constraints already known and discussed in the literature \cite{Iliesiu:2018fao, Alday:2020eua}, with the goal of formulating a complete bootstrap problem. 
    \paragraph{KMS condition} The periodicity of correlations functions for finite temperature theories is known in the literature as KMS condition \cite{Kubo:1957mj,Martin:1959jp,Iliesiu:2018fao, Alday:2020eua}. This condition reads \begin{equation}\label{KSM condition}
        f_{\beta}(\tau,|\vec x|) = f_{\beta}(\tau+\beta,|\vec x|) \ .
    \end{equation}
    \paragraph{Analyticity} Let us define in the complex plane the variables $\omega = \text{Arg}(\tau+i |\vec x|)$, and $r = \text{Abs}(\tau+i |\vec x|)$. In these coordinates, the two-point function reads
    \begin{equation}
        f_{\beta}(\tau,|\vec x|)=f_{\beta}\left(\frac{r}{\omega}, r \omega \right).
    \end{equation}
    Then, the analiticity consistency condition states that the two-point function can only have cuts starting in $\omega = \pm r$ and $\omega = \pm r^{-1/2}$; moreover, it has poles in $\omega = 0, \pm \tilde \omega$, where  $\tilde \omega = \sqrt{(\tau+i|\vec x|)/(\tau-i|\vec x|)}$. Outside of these poles and cuts, the two-point function must be an analytic function \cite{Iliesiu:2018fao}. This consistency condition is illustrated in Figure \ref{Analiticity}.
    \paragraph{Regge limit} Let us define in the complex plane the variables $z=\tau+i|\Vec{x}|$ and $\Bar{z}=\tau-i|\Vec{x}|$. With this set of coordinates, we can introduce the variables $\rho= \sqrt{z \overline z}$ and $\eta = \frac{z+\overline z}{2 \sqrt{z\overline z}}$. In these coordinates, the two-point function reads $f_{\beta}\left(\rho, \eta \right)$. The Regge limit is
    \begin{equation}
        \lim_{|\eta| \rightarrow \infty} f_{\beta}\left(\rho, \eta \right) \ , \quad \rho \text{ fixed} \ .
    \end{equation}
    The two-point function must be polynomially bounded in the Regge limit \cite{Iliesiu:2018fao,Alday:2020eua}\footnote{Analiticity in the spin is a consequence of polynomial boundedness in the $|\eta |\to \infty$ limit, since this implies that the two-point function does not grow faster than $\eta^{J_\text{Regge.}}$. This is the analog of the Regge limit at zero temperature \cite{Iliesiu:2018fao,Alday:2020eua}.}.
    \paragraph{Large distance limit} As mentioned before, it is believed that the compactification on the thermal manifold produces a massive theory with a mass gap $m_{\text{th}}>0$ \footnote{By dimensional analysis this mass has to be proportional to $\beta^{-1}$. Moreover, notice that, as already suggested in \cite{Iliesiu:2018fao, Benjamin:2023qsc}, the existence of a thermal mass is not always guaranteed: in some cases (e.g. bosonic free theory) symmetries preserve the theory from a mass gap. This could also be the case if the theory spontaneously breaks a symmetry at finite temperature as in the case of \cite{Chai:2020onq,Chai:2020zgq}, however, we are not aware of any example in integer spacetime dimension. Nevertheless, in a generic interacting CFT at finite temperature the existence of such a mass gap is expected from the fact that thermal fluctuations make correlations between local operators negligible at large distances.}. Assuming this folk-theorem, the two-point function $f_{\beta}(\tau,|\vec x|)$ approaches spatial infinity as \cite{Iliesiu:2018fao,Caron-Huot:2022akb}\begin{equation}\label{eq: clustering}
        f_{\beta}(\tau,|\vec x|) \sim \Braket{\mathcal{O}_{1}}_\beta  \Braket{\mathcal{O}_{2}}_\beta+ \mathcal O \left (e^{-m_{\text{th}}|\vec x|}\right) \ .
    \end{equation}
    \paragraph{Zero temperature limit} Since a thermal two-point function can be defined on a thermal manifold with an arbitrary value of $\beta$, we expect $f_{\beta}\left(\tau, |\Vec{x}|\right)$ to be a continuous function in $\beta$; moreover, the physical expectation is for it to be smooth and to have a well-defined limit for $\beta \rightarrow \infty$
    \begin{equation}\label{LargeBetaLimit}
       \lim_{\beta \rightarrow \infty} \Braket{\mathcal{O}_1(x_1) \mathcal{O}_2(x_2)}_{\beta}=\Braket{\mathcal{O}_1(x_1) \mathcal{O}_2(x_2)}_{\mathbb{R}^{d}} \ .
    \end{equation}
    \paragraph{OPE}  When two local scalar operators $\mathcal{O}_1$ and $\mathcal{O}_2$, with conformal dimensions $\Delta_1$ and $\Delta_2$ respectively, lie in a sphere whose interior is flat, the OPE is exact and reads \cite{Iliesiu:2018fao}
    \begin{equation}
        \co_{1}(x_1) \times \co_{2}(x_2) \sim \sum_{\mathcal O \in \mathcal{O}_1 \times \mathcal{O}_2} \frac{f_{\mathcal{O}_1\mathcal{O}_2 \mathcal O}}{c_{\mathcal O}}\left(\tau^2+|\vec x|^2\right)^{\frac{1}{2}\left(\Delta_{\co}-\Delta_1 - \Delta_2-J\right)}x_{\mu_1}\ldots x_{\mu_J} \mathcal O^{\mu_1\ldots \mu_J}(x_2) \ ,
    \end{equation}
    where $x^{\mu}=x_{1}^{\mu}-x_{2}^{\mu}$, $\Delta_{\co}$ is the conformal dimensions of $\mathcal{O}$, $f_{\mathcal{O}_1\mathcal{O}_2 \mathcal O}$ is the three-point function OPE constant at zero temperature, and $c_{\mathcal O}$ is the normalization coefficient appearing in the two-point function at zero temperature \begin{equation}
        \langle \mathcal O^{\mu_1\ldots \mu_J}(x) \mathcal O_{\nu_1\ldots \nu_J}(0)\rangle = c_{\mathcal O} \frac{I_{(\nu_1\ldots \nu_J)}^{\,(\mu_1\ldots \mu_J)}}{|x|^{2\Delta_{\co}}} \ , \hspace{1 cm}I_{\, \nu}^{\mu} = \delta_{\, \nu}^{\mu}- \frac{2 x_\nu x^\mu}{x^2}\ .
    \end{equation}
   The geometry of the thermal manifold imposes a minimal radius of convergence to the OPE since the largest radius for a sphere with a flat interior is $\beta$. Consistency with the OPE regime requires the thermal two-point function to read, inside the convergence radius, 
    \begin{equation}
        f_{\beta}(\tau,|\vec x|) = \sum_{\mathcal O \in \mathcal{O}_1 \times \mathcal{O}_2} \frac{f_{\mathcal{O}_1\mathcal{O}_2 \mathcal O}}{c_{\mathcal O}}\left(\tau^2+|\vec x|^2\right)^{\frac{1}{2}\left(\Delta_{\co}-\Delta_1 - \Delta_2-J\right)}x_{\mu_1}\ldots x_{\mu_J} \Braket{\mathcal O^{\mu_1\ldots \mu_J}}_{\beta} \ . \label{eq: OPE cond} 
    \end{equation}
    Working in the OPE regime, it is possible to apply the constraints derived from the broken and unbroken Ward identities to the expression \eqref{eq: OPE cond}:  we will explore this in the Section \ref{subsec: constraints in the OPE regime}. At this stage let us comment that the two-point function has to be consistent with the OPE decomposition and therefore the anomalous dimensions are those at zero temperature \cite{Alday:2020eua}.
  \newline  The bootstrap problem for the thermal scalar two-point function $\Braket{\co_{1}(x_1)\co_2(x_2)}_{\beta}$ consists of finding a function of the coordinates $x_1$, $x_2$ and the inverse of the temperature $\beta$  such that it satisfies all the conditions listed in this Section. First of all, the unbroken Ward identities set the dependence on the coordinates to be like in Equation \eqref{eq: coord dep}. Then:
    \begin{enumerate}
        \item it must be periodic over the thermal circle (cfr. Equation \eqref{KSM condition});
        \item in the $\omega$ plane, it must be analytic everywhere apart from the poles and cuts illustrated in Figure \ref{Analiticity};
    \item its Regge limit must be polynomially bounded; its zero temperature limit must return the same two-point function, but computed on the flat space; the large distance limit must return a clustered structure like in Equation \eqref{eq: clustering};
    \item it must be compatible with the OPE expansion in the OPE regime, i.e. $\sqrt{\tau^2+\vec x^2} \leq \beta$;
    \item it must be symmetric with respect to $\tau \to - \tau$. Following \cite{Iliesiu:2018fao}, if the CFT at zero temperature is not parity invariant under the transformation $\tau \to -\tau$, then this transformation has to be performed together with a reflection of one (or more) direction in the spatial component $\mathbb R^{d-1}$. Both realizations of the parity invariance enhance the symmetry group preserved at finite temperature using a discrete, additional symmetry;
    \item it must satisfy the constraints \eqref{eq: rot2 2pts funct} and \eqref{eq: dil 2pts funct}, imposed by the broken symmetries of the Global Conformal Group.
    \end{enumerate}
    The last point of this list represents the novelty of this work with respect to the already existing literature on the subject; nevertheless, we can also think of these new conditions as additional constraints to bootstrapping one-point functions at the finite temperature given the two-point functions (including the stress-energy tensor) or, even more intriguingly, as a consistency condition for the zero-temperature conformal field theory. As already anticipated, in Section \ref{subsec: constraints in the OPE regime} we will combine the OPE expansion with these newly found constraints. 
    \newline
    \newline If the bootstrap problem set above returns unique quantities or not is an unsolved problem; in particular, the minimal amount of conditions to impose to uniquely determine the two-point functions is not known. Suppose the existence of two different solutions $f^{1}$ and $f^{2}$ to the problem. Imagine first that the two solutions differ by the contribution of a finite number of local operators in the OPE regime: then, the large distance limit of the difference $\Delta f = f_1-f_2$ is\footnote{The limit holds only for even spin operators; however it is sufficient because the one point function of an odd spin operator is zero and their contribution to the two-point function thereof.} \begin{equation}
       \lim_{r\to \infty}\left | \Delta f(\tau,r)\right | =  \infty  \ ,
   \end{equation}
   and therefore if $f_1$ satisfies the clustering property, expressed in equation \eqref{eq: clustering}, then $f_2$ does not satisfy it (and \emph{vice-versa}), so only one among $f_1$ and $f_2$ is a solution to the bootstrap problem.  If $f_1$ and $f_2$ differ for an overall factor then the limit \eqref{LargeBetaLimit} differ for the same overall factor, which can be identified with the normalization of the two-point function at zero temperature. If we fix this normalization before solving the bootstrap problem, then it fixes the residue of the poles of the two-point function at finite temperature.
   Two solutions $f_1$ and $f_2$ that do not differ by an overall factor must have the same poles with the same residues since the poles are given by the zero temperature contribution and the limit $\beta \to \infty$ defines the coefficients in front.
   \newline We couldn't find a way to discriminate between two solutions complying with all the reasoning above but differing for one or more branch cuts. This problem was already pointed out in \cite{Alday:2020eua} where the authors proved unicity assuming some specific conditions\footnote{The necessary condition for the unicity proposed in  \cite{Alday:2020eua} is $\text{Disc}(z,\overline z) = 0$, $0 \le \overline z = \tau-i r \le 1$ and $1 \le z = \tau+i r$.}, by using the Regge boundedness of the thermal correlator and the subtracted thermal dispersion relations.
   Observe that, however, in the case of a free scalar theory in four spacetime dimensions, it is possible to reconstruct the thermal two-point function of the free scalar theory in four dimensions by only using the bootstrap conditions recapped above (cfr. Appendix \ref{sec: Free Scalar Tests of the Broken Ward Identities}) . This method seems to work only in the case of the free scalar and the deep reason for that is the simplicity of the OPE expansion between two fundamental scalars that allows for a factorization of the two-point function in terms of \emph{holomorphic} functions. In general, it is difficult to read this factorization from the OPE (and we do not expect that every two-point function can be factorized in a similar way), however, we do not exclude that a more general unicity can be proven with more sophisticated tools. We leave this to future studies.
 \subsection{Broken Ward identities in the OPE regime}\label{subsec: constraints in the OPE regime}
 The constraints set up by the broken symmetries of the global Conformal Group are difficult to solve by themselves since the structure of the thermal breaking term requires the knowledge of a thermal $(n+1)$-point function in order to study a thermal $n$-point function. However, in this Section we show how the breaking term splits into lower-point functions in the OPE regime, allowing for an explicit derivation of the solution to the bootstrap problem and for additional constraints on the thermal one-point functions. We are going to focus on (broken) dilatations and rotations; in the Appendix \ref{Susy in OPE}, we show instead that a generic broken Ward identity, when combined with the OPE, always returns the Leibniz rule for the generators of the broken symmetry. 
 \newline Let us start by focusing on the new constraint coming from the dilatations broken Ward identity. We study the thermal scalar two-point function 
    \begin{equation}
        f_{\beta}(\tau, |\Vec{x}|)=\Braket{\phu(x)\phu(0)}_{\beta} \ ,
    \end{equation}
    where both scalar operators now have conformal dimension $\Delta_{\phu}$. By introducing the radial coordinate $r=| \Vec{x}|$, the dilatations broken Ward identity \eqref{eq: dil 2pts funct} reads
	\begin{equation} \label{eq: dil br wi to solve}
		\left(\tau \frac{\partial}{\partial \tau}+ r \frac{\partial}{\partial r}+2\Delta_\phu  \right) f_{\beta}\left(\tau,  r\right) =\beta \int d^{d-1}y \Braket{T^{00}(0,\vec y)\phu(x)\phu(0)}_{\beta} \ . 
	\end{equation}
	If we consider the operators $\phu$ in the OPE regime, the broken Ward identity can be rewritten as
	\begin{multline}
		\left(\tau \frac{\partial}{\partial \tau}+ r \frac{\partial}{\partial r}+2\Delta_\varphi  \right) f_{\beta}\left(\tau,  r\right) = \\= \beta \sum_{\co \in \phu \times \phu}\frac{f_{\phu\phu\co}}{c_{\co}} \left(\tau^2+ r^2  \right)^{\frac{1}{2}\left(\Delta_{\co}-2\Delta_\varphi-J \right)} x_{\mu_{1}}\dots x_{\mu_{J}}\int d^{d-1}y \Braket{T^{00}(0,\vec y)\co^{\mu_{1}\dots \mu_{J}}\left(0\right)}_{\beta}\ .
	\end{multline}
	The right-hand side can be rewritten by plugging the equation \eqref{eq: 1pt thermal function}, which is the dilatations broken Ward identity applied to thermal one-point functions
	\begin{multline}\label{eq: Differential equations}
		\left(\tau \frac{\partial}{\partial \tau}+ r \frac{\partial}{\partial r}+2\Delta_\varphi \right) f_{\beta}\left(\tau,  r\right) = \\= \sum_{\co \in \phu \times \phu}\frac{f_{\phu\phu\co} \Delta_{\co}}{c_{\co}} \left(\tau^2+ r^2  \right)^{\frac{1}{2}\left(\Delta_{\co}-2\Delta_\varphi-J \right)} x_{\mu_{1}}\dots x_{\mu_{J}} \Braket{\co^{\mu_{1}\dots \mu_{J}}}_{\beta} \ .
	\end{multline}
The left-hand side of the equation \eqref{eq: Differential equations} respects the OPE regime as well, so it can be expanded in a similar fashion. We get
\begin{equation}
		 \sum_{\co \in \phu \times \phu}\left(\tau \frac{\partial}{\partial \tau}+ r \frac{\partial}{\partial r}+2\Delta_\varphi- \Delta_{\co}\right)f_\beta^{(\mathcal O)}(\tau, r)=0 \ ,
	\end{equation}
where 
\begin{equation}\label{eq: dil bwi sum}
     f_\beta^{(\mathcal O)}(\tau, r)= \frac{f_{\phu\phu\co}}{c_{\co}} \left(\tau^2+ r^2  \right)^{\frac{1}{2}\left(\Delta_{\co}-2\Delta_\varphi-J \right)} x_{\mu_{1}}\dots x_{\mu_{J}} \Braket{\co^{\mu_{1}\dots \mu_{J}}}_{\beta} \ . 
 \end{equation}
Since $\langle \mathcal O^{\mu_1\ldots \mu_J}\rangle_\beta \propto \beta^{-\Delta_{\mathcal O}}$ by dimensional analysis, we can interpret the sum over the operators in the OPE appearing in the equation \eqref{eq: dil bwi sum} as a sum over powers of $\beta$\footnote{This requires additional care in the case of a degenerate spectrum. In the latter case, we can either distinguish the two operators from the spin or from other global symmetries; in the first case, the contribution of any operator is always captured by equation \eqref{eq: eq1 comp} by the linearity of the differential equation \eqref{eq: dil bwi sum}, in the second case the one-point function is zero because the operator is not in the trivial representation of the global symmetry's group.}. Hence, each term of the series constitutes an independent differential equation
 \begin{equation}\label{eq: inhomogeneous equations}
    \left(\tau \frac{\partial}{\partial \tau}+ r \frac{\partial}{\partial r}+2\Delta_\varphi \right) f_\beta^{(\mathcal O)}(r,\tau) = \Delta_{\mathcal O}  f_\beta^{(\mathcal O)} \ .
 \end{equation}
 If we consider the identity operator $\mathds{1} \in \co_1 \times \co_2$, then $\Delta_{\mathds{1}}=0$ and the equation \eqref{eq: inhomogeneous equations} returns the zero temperature result. For all the other cases we can write down also the boosts broken Ward identity \begin{equation} 
		\left( \tau r \frac{\partial}{\partial r}-r^2 \frac{\partial}{\partial \tau}\right)  f_{\beta}\left(\tau,  r \right)=x_{i}\int d^{d-1}y \Braket{T^{0i}(0,\vec y)\phu(x) \phu(0)}_{\beta} \ .
	\end{equation}
	This equation can be expanded in OPE as well; we expand the right-hand side first 
	\begin{multline}
		\left( \tau r \frac{\partial}{\partial r}-r^2 \frac{\partial}{\partial \tau}\right)  f_{\beta}\left(\tau,  r \right)=\\=\beta \sum_{\co \in \phu \times \phu}\frac{f_{\phu\phu\co}}{c_{\co}} \left(\tau^2+ r^2  \right)^{\frac{1}{2}\left(\Delta_{\co}-2\Delta_\varphi-J \right)} x_{\mu_{1}}\dots x_{\mu_{J}}x_{i}\int d^{d-1}y \Braket{T^{0i}(0,\vec y)\co^{\mu_{1}\dots \mu_{J}}\left(0\right)}_{\beta} \ ,  
	\end{multline}
 and then we apply the broken Ward identity \eqref{eq: 1pt functions rot} to rewrite the integral as
 \begin{multline} \label{eq: rightHandside}
		\left( \tau r \frac{\partial}{\partial r}-r^2 \frac{\partial}{\partial \tau}\right)  f_{\beta}\left(\tau,  r \right)=\\=i\beta  \sum_{\co \in \phu \times \phu}\frac{f_{\phu\phu\co}}{c_{\co}} \left(\tau^2+ r^2  \right)^{\frac{1}{2}\left(\Delta_{\co}-2\Delta_\varphi-J \right)} x_{\mu_{1}}\dots x_{\mu_{J}}x_{i} \Braket{\op S_{i0}\co^{\mu_{1}\dots \mu_{J}}\left(0\right)}_{\beta} \ . 
	\end{multline}
 The equation \eqref{eq: inhomogeneous equations} together with the equation \eqref{eq: rightHandside} can be interpreted as a system of partial differential equations. In general, we have that the first differential equation, i.e. \eqref{eq: inhomogeneous equations}, fixes  \begin{equation}
     f_\beta^{\mathcal O}(\tau,r) \propto \left (r^2+\tau^2\right )^{\frac{1}{2}\left(\Delta_{\mathcal O}-2\Delta_{\varphi}-J\right )}\tau^J g_{\mathcal O}\left(\frac{r}{\tau}\right) \ ,
 \end{equation}
 where the function $g_{\mathcal O}(r/\tau)$ remains unfixed. The next step is to plug this solution in the rotation broken Ward identities; order by order in the OPE expansion equation \eqref{eq: rightHandside} reads \begin{equation}
     \frac{r}{\tau^2} \tau^J \left(-J r \tau g_{\mathcal O}\left(\frac{r}{\tau}\right)+\left(r^2+\tau^2\right)g_{\mathcal O}'\left(\frac{r}{\tau}\right)\right) = x_{\mu_1}\ldots x_{\mu_J}x_i \left \langle \op S_{i0}\mathcal O^{\mu_1\ldots \mu_J} \right \rangle_\beta \ .
 \end{equation}
 For scalars, i.e. $J = 0$, we get $g_{\mathcal O}(r/\tau) = \text{const}.$, meaning that the one-point function of a scalar $\left \langle \mathcal O \right \rangle_\beta $ does not have any kinematical structure, as expected. For vectors, i.e. $J = 1$,  by imposing the reality condition on the solution\footnote{The differential equation is also solved by a second function proportional to a hypergeometric function: however, this function is not real for all the possible values of $r$ and $\tau$.}, we get $\left \langle \mathcal O^\mu \right \rangle = 0$ and only a term proportional to an odd power of $\tau$ can contribute to the OPE. However, this contribution is zero because of the symmetry $\tau \to - \tau$, which forbids odd functions of $\tau$ to appear\footnote{The use of $\tau \to -\tau$ is necessary to get the correct result. This corresponds to the use of the full $\mathrm O(d-1)$ symmetry group in \cite{Iliesiu:2018fao}: $\mathrm{SO}(d-1)$ is indeed not sufficient to exclude odd spin operators to have a non-zero thermal one-point function.}. For 2-rank tensors, i.e. $J=2$, we can recall the results of Section \eqref{sec Two rank operators} to see that the solution to the two differential equations is given by 
 \begin{equation}
     f_\beta^{(\mathcal O)}(\tau,r) \propto \left(\tau^2+ r^2  \right)^{\frac{1}{2}\left(\Delta_{\co}-2\Delta_\varphi \right)} C_{2}^{(\nu)}\left(\frac{\tau}{\sqrt{\tau^2+r^2}}\right ) \ .\label{eq: eq1 comp}
 \end{equation}
 This result can be extended to all values of $J$ by observing that the first differential equation is solved by the ansatz
 \begin{equation}
     f_\beta^{(\mathcal O)}(\tau,r) = \frac{\mathcal{C}_{J,\nu}^{\mathcal{O}}}{\beta^{\Delta_{\co}}}\left(\tau^2+ r^2  \right)^{\frac{1}{2}\left(\Delta_{\co}-2\Delta_\varphi \right)} C_{J}^{(\nu)}\left(\frac{\tau}{\sqrt{\tau^2+r^2}}\right ) \ , \label{eq: eq1 comp}
 \end{equation}
 where the $\beta^{-\Delta_{\co}}$ factor is introduced by dimensional analysis and $\mathcal{C}_{J,\nu}^{\mathcal{O}}$ is an overall coefficient to be fixed. We can fix it by imposing consistency of the thermal two-point function with the OPE. By comparing the expressions  \eqref{eq: dil bwi sum} and \eqref{eq: eq1 comp} we identify
 \begin{equation}
     \frac{\mathcal{C}_{J,\nu}^{\mathcal{O}}}{\beta^{\Delta_{\co}}} C_{J}^{(\nu)}\left(\frac{\tau}{\sqrt{\tau^2+r^2}}\right )=\frac{f_{\phu\phu\co}}{c_{\co}} \left(\tau^2+ r^2  \right)^{-\frac{J}{2}} x_{\mu_{1}}\dots x_{\mu_{J}} \Braket{\co^{\mu_{1}\dots \mu_{J}}}_{\beta} \ .
 \end{equation}
 If $\nu=\frac{d-2}{2}$, the following identity holds \cite{Iliesiu:2018fao}
 \begin{equation}
     C_J^{(\nu)}\left(\frac{\tau}{\sqrt{\tau^2+r^2}}\right) = \frac{2^J(\nu)_J}{J!}\left(\tau^2+ r^2  \right)^{-\frac{J}{2}} x_{\mu_1}\ldots x_{\mu_J}\left(e^{\mu_1}\ldots e^{\mu_J}-\text{traces}\right) \ ,
 \end{equation}
 leading to 
 \begin{equation}
     \frac{\mathcal{C}_{J,\nu}^{\mathcal{O}}}{b_{\co}}\frac{2^J(\nu)_J}{J!} \frac{b_{\co}}{\beta^{\Delta_{\co}}}\left(e^{\mu_1}\ldots e^{\mu_J}-\text{traces}\right)=\frac{f_{\phu\phu\co}}{c_{\co}}  \Braket{\co^{\mu_{1}\dots \mu_{J}}}_{\beta} \ ,
 \end{equation}
 from which we can both recover the formula \eqref{eq: one point functions} and fix the overall coefficient
\begin{equation}
       f_\beta^{(\mathcal O)}(\tau,r) = \frac{J!}{2^J (\nu)_J}\frac{1}{\beta^{\Delta_{\co}}}\frac{f_{\phu\phu\co} b_{\co}}{c_{\co}} \left(\tau^2+ r^2  \right)^{\frac{1}{2}\left(\Delta_{\co}-2\Delta_\varphi \right)} C_{J}^{(\nu)}\left(\frac{\tau}{\sqrt{\tau^2+r^2}}\right ) \ ,
 \end{equation}
 which immediately leads to the final solution\footnote{The procedure highlighted in this short Section shows how it is possible to recover the results already presented in \cite{Iliesiu:2018fao} by employing only the broken Ward identities and the OPE. It must be noticed that strictly speaking, the formula \eqref{eq: one point functions} was not recovered in full generality, but only for those operators living in the OPE between two identical scalar operators; moreover, the second differential equation has to be solved spin by spin.}
 \begin{equation}\label{eq: OPE decomposition}
       f_\beta(\tau,r) = \sum_{\mathcal{O}\in \phu \times \phu}\frac{J!}{2^J (\nu)_J}\frac{1}{\beta^{\Delta_{\co}}}\frac{f_{\phu\phu\co} b_{\co}}{c_{\co}} \left(\tau^2+ r^2  \right)^{\frac{1}{2}\left(\Delta_{\co}-2\Delta_\varphi \right)} C_{J}^{(\nu)}\left(\frac{\tau}{\sqrt{\tau^2+r^2}}\right ) \ .
 \end{equation}
 Further, observe that the OPE decomposition of the two-point function implies that odd spin operators' one-point functions have to be zero; this can be easily checked by considering the fact that $G_{J}^{\nu}(\tau/\sqrt{r^2+\tau^2}) = G_{J}^{\nu}(-\tau/\sqrt{r^2+\tau^2})$ only for even spin operator and therefore $\tau \to - \tau$ is a symmetry only if the one-point functions of odd spin operator are zero. 
 \newline In conclusion, working with the broken Ward identities in the OPE regime allowed us to obtain already-known results about the structure of the one-point functions and the OPE decomposition of the two-point functions. This was expected since broken Ward identities contain symmetry information about broken and unbroken symmetry. These results also suggest that broken Ward identities could contain more information, possibly also outside the OPE regime.
\section{Superconformal broken Ward identities}\label{sec: Superconformal BWI}
Whether any trace of supersymmetry is preserved at finite temperature has been a longstanding question  \cite{Fuchs:1984ed,Aoyama:1984bk,Das:1978rx,Girardello:1980vv}, and significant progress has been made recently with important developments and applications \cite{Friess:2006fk,Berenstein:2022nlj,Caron-Huot:2022akb,Caron-Huot:2022lff,Alday:2020eua,Petkou:2021zhg,Karlsson:2019qfi,Karlsson:2021duj,Karlsson:2022osn,Benjamin:2023qsc}. The discussion on the broken Ward identities (at finite temperature) presented in Section \ref{sec: Broken Ward Identities at Finite Temperature} can be specialized to the case of superconformal field theories.
In the following, we will focus on supersymmetric and superconformal theories in four dimensions; however, the arguments and the procedures can be adapted to other dimensions.
\newline The Superconformal Group extends the Conformal Group with supersymmetry generators $\op Q_\alpha^{I}$ (and $\op {\overline Q}^{\dot \alpha I}$), superconformal generators $\op S_\alpha^I$ (and $\op {\overline S}^{\dot \alpha I}$) and R-symmetry generators $\op R^{IJ}$ satisfying precise commutator relations producing specific decomposition of the Hilbert space of the theory \cite{Eberhardt:2020cxo,Dolan:2001tt,Dolan:2002zh}.
 \newline Due to the anti-periodicity of the fermionic operators over the thermal circle $S^1_{\beta}$, we expect supersymmetry to be broken; however, it is not obvious \emph{a priori} if the R-symmetry is broken or unbroken. We derive in this Section explicit broken Ward identities for all the superconformal generators.

 Since the discussion in this Section is completely analogous to the one presented in Section \ref{sec: Broken Ward Identities at Finite Temperature} we do not derive again step by step the (broken) Ward identities, but we will simply discuss the differences between conformal generators and superconformal ones.
 \subsection{Supersymmetry and superconformal symmetry}
 The procedure underlined in Section \ref{sec: Broken Ward Identities at Finite Temperature} can be followed in every detail, with a few remarks. In particular, the first difference concerns the variation of the action with respect to a symmetry with fermionic generators\footnote{In the following, the spinor indices $\alpha, \beta, \gamma$ will run over $\lbrace 1,2\rbrace $; the same thing applies to the dotted spinor indices $\dot{\alpha}, \dot{\beta}, \dot{\gamma}$. The R-symmetry indices will be denoted by the set $I,J,K$ and will run over $\lbrace 1, \dots, \mathcal{N} \rbrace$, with $\mathcal{N}$ representing the amount of supersymmetry.}
 \begin{equation}
     \delta S = \int_{\mathcal M} d^d x \sqrt{g} \ \nabla_\mu \xi^\alpha(x) \ G_{\alpha}^{\mu}(x) \ ,
 \end{equation}
 where $G_{\alpha}^{\mu}$ is the fermionic current, $g_{\mu \nu}$ is the metric on the manifold $\mathcal{M}$ and $\xi^{\alpha}$ is an infinitesimal spinor. From the above equation it is clear that, on a generic non-flat manifold, all the fermionic generators are broken since there is not a covariantly constant spinor. At finite temperature, i.e. when $\mathcal M_\beta =\mathbb R^{d-1}\times  S^1_{\beta} $, the spacetime is locally flat and this argument does not prevent fermionic generators from realizing quantum symmetries. We can then proceed straight-forwardly as in the bosonic case. A second remark can be observed in the evaluation of the following spatial boundary integration
 \begin{equation}
    \lim_{R\to \infty} \left \langle J_a^{i}(\tau,R,\Omega) \mathcal O_1(x_1) \ldots \mathcal O_n(x_n) \right \rangle_\beta = \lim_{R \to \infty }\left \langle J_a^{i}(\tau,R,\Omega)\right \rangle_\beta \left \langle\mathcal O_1(x_1) \ldots \mathcal O_n(x_n) \right  \rangle_\beta = 0 \ .
 \end{equation}
 Differently from the bosonic case, this contribution does not return an infrared divergence because the one-point function of the fermionic current is necessarily zero: this could follow from the solution of the rotation broken Ward identity, or from a symmetry argument \cite{Iliesiu:2018fao}. The rest of the procedure to compute broken Ward identities is not altered. In order to discuss the breaking terms it is useful to write down the \textit{superconformal current} 
 \begin{equation}\label{eq: superconformal currents}
     J_{SC}^\mu(x) = \psi_I^\alpha(x) G_\alpha^{\mu I }(x)\ ,
 \end{equation}
  where $\psi_I^\alpha(x)$ is the conformal Killing spinor  \begin{equation}
     \psi_{I}^\alpha (x) = \lambda_I^\alpha + x_\nu \left(\overline \sigma^{\nu }\right)^{\, \alpha \dot \alpha} \mu_{I \dot \alpha}  \ ,
 \end{equation}
 where $\mu_{I \dot \alpha}$ and $\lambda_I$ are constant spinors. 
At zero temperature, we define the supercharges and the superconformal charges\footnote{As it was done in the example \eqref{eq: dil example}, the correct definition of the quantum superconformal generators must take into account the correct spacetime dependence $$\op {\overline S^{\dot \alpha} } \, \mathcal O(x) = \int_{\mathbb R^{d-1}}d^{d-1} y \ \left(\overline \sigma^{\nu }\right)_{\, \alpha \dot \alpha}(y_\nu-x_\nu) \ G^{0\alpha}(y) \mathcal O(x) \ .$$ } \begin{equation}
     Q^\alpha = \int_{\mathbb R^{d-1}}d^{d-1} x \ G^{0\alpha}(x) \ , \hspace{1 cm} \overline S^{\dot \alpha} = \int_{\mathbb R^{d-1}}d^{d-1} x \  \left(\overline \sigma^{\nu }\right)_{\, \alpha \dot \alpha}x_\nu \ G^{0\alpha}(x) \ ,
 \end{equation}
When $\lambda_I^{\alpha} = 0$, $\mu_{I \dot \alpha} \neq 0$ the current \eqref{eq: superconformal currents} corresponds to the symmetry generated by superconformal generators; instead when $\mu_{I \dot \alpha} = 0$, $\lambda_I^{\alpha} \neq 0$  the current is associated to supersymmetry generators.
 \newline Due to the fermionic nature of $G_{\alpha}^{\mu I}(\tau,\vec x)$, and the anti-periodicity of fermionic operators with respect to the thermal circle, it is natural to write down the most general breaking term, generally defined in \eqref{eq: breaking term definition}, for the superconformal current \eqref{eq: superconformal currents}, as \begin{equation}
      \left[2 \lambda_I^{\alpha}+\left((\beta-2\tau) \left(\overline \sigma_0\right)^{\alpha \dot \alpha}+2(x_i-y_i)\left(\overline \sigma^i\right)^{\alpha \dot \alpha}\right)\mu_{I \dot \alpha}\right] G_{\alpha }^{0 I}(\beta,\vec y) \ ,
 \end{equation}
 where the coordinates $(\tau,\vec y)$ are the coordinates of the operator to which the generator is applied.
 The expression above agrees with the physical expectation according to which the fermionic nature of $\op Q_\alpha^I$ and $\op S_\alpha^I$ implies an explicit breaking of supersymmetry at finite temperature. Leaving the spinor indices and the R-symmetry indices implicit for the sake of notation clarity, the broken Ward identities read
 \begin{equation}\label{eq: SBWI}
     \sum_{r=1}^{n} \left \langle \mathcal O_1(x_1) \ldots \left[\op Q,\mathcal O_r\right ] (x_r)\ldots \mathcal O_n(x_n) \right \rangle_\beta = 2 \int d^{d-1} y \ \left \langle G^{0}(\beta ,\vec y)\mathcal O_1(x_1) \ldots \mathcal O_n(x_n) \right \rangle_\beta \ ,
 \end{equation}
 \begin{multline}
     \sum_{r=1}^{n} \left \langle \mathcal O_1(x_1) \ldots \left [\Bar {\op S},\mathcal O_r\right ](x_r)\ldots \mathcal O_n(x_n) \right \rangle_\beta =\\= \sum_{r=1}^{n}(\beta-2 \tau_r)\overline \sigma^0 \int d^{d-1}y \left \langle G^{0}(\beta ,\vec y)\mathcal O_1(x_1) \ldots \mathcal O_n(x_n) \right \rangle_\beta + \\+  \sum_{r=1}^{n}\overline \sigma^j \int d^{d-1}y \ (y_j-x_j)\left \langle G^{0}(\beta ,\vec y)\mathcal O_1(x_1) \ldots \mathcal O_n(x_n) \right \rangle_\beta    \  .
 \end{multline}
 These two equations encode the superconformal symmetry breaking at finite temperature. Equation \eqref{eq: SBWI} rules the behaviour of correlation functions on which we act with the supersymmetry generators. In particular the breaking term contains a correlation function in which the supercurrent appear. In deriving those equations it is completely clear that the fact that supersymmetry is broken at finite temperature is therefore due to the anti-periodic boundary conditions given to fermions. For completeness, we also write down the broken Ward identities for right chiral supersymmetry generators \begin{equation}\label{eq: SBWI2}
     \sum_{r=1}^{n} \left \langle \mathcal O_1(x_1) \ldots \left [\Bar{\op Q},\mathcal O_r \right](x_r)\ldots \mathcal O_n(x_n) \right \rangle_\beta = 2 \int d^{d-1} y \ \left \langle \Bar{G}^{0}(\beta ,\vec y)\mathcal O_1(x_1) \ldots \mathcal O_n(x_n) \right \rangle_\beta \ ,
 \end{equation}
 \begin{multline}
     \sum_{r=1}^{n} \left \langle \mathcal O_1(x_1) \ldots [ {\op S},\mathcal O_r](x_r)\ldots \mathcal O_n(x_n) \right \rangle_\beta =\\ =\sum_{r=1}^{n}(\beta-2 \tau_r) \sigma^{0}\int d^{d-1}y \left \langle \Bar{G}^{0}(\beta ,\vec y)\mathcal O_1(x_1) \ldots \mathcal O_n(x_n) \right \rangle_\beta + \\+ \sum_{r=1}^{n} \sigma^j\int d^{d-1}y \ (y_j-x_j) \left \langle \Bar{G}^{0}(\beta ,\vec y)\mathcal O_1(x_1) \ldots \mathcal O_n(x_n) \right \rangle_\beta    \  \ ,
 \end{multline}
 \subsection{The fate of R-symmetry at finite temperature}
 The Superconformal Group includes the R-symmetry which can be understood as a rotation of the supercharges. Even if supersymmetry is broken at finite temperature it is not immediately clear what happens to the R-symmetry. To address this question and demonstrate the power and generality of the discussion in Section \ref{sec: Broken Ward Identities at Finite Temperature}, let us explicitly compute the breaking term that appears in the (broken) Ward identities. In order to do that, we need to recall that the R-symmetry generators can be defined as
 \begin{equation}
      R^{IJ} = \int_{\mathbb R^{d-1}} d^{d-1}x \ J^{0 , IJ}(x) \ ,
 \end{equation}
 where $J^{\mu, IJ}$ is a bosonic current. Therefore we must consider the breaking term defined by the equation \eqref{eq: breaking term definition} in the case of the R-symmetry current. Since there is no additional coordinate dependence (as in the case of the dilatation broken Ward identity, for example) and the current is bosonic (so it is periodic over the thermal circle), we conclude that the breaking term is simply zero, the Ward identity is unbroken and R-symmetry is preserved.
 \newline This result is noteworthy because it shows that in a  supersymmetric/superconformal field theory at finite temperature, although the  supersymmetry/superconformal symmetry is broken, the R-symmetry remains intact as a global symmetry acting on the operators of the theory. This has important consequences, as it implies that every operator charged under the Abelian $\mathfrak{u}(1)$ sub-algebra of the R-symmetry at zero temperature must have a zero one-point function: in formula \begin{equation}
     \left \langle \mathcal O \right \rangle_\beta = 0 \ , \hspace{1 cm}\text{charged} \  . 
 \end{equation}
 For instance, in the case of the $\mathcal N = 4$ SYM theory, this means that in the stress-tensor multiplet there are only two operators that are allowed to have a non-zero one-point function: the Lagrangian operator and the stress-energy tensor. All the other operators are charged under R-symmetry and therefore their thermal vacuum expectation values are zero.

  Another way to see that the R-symmetry is preserved is to consider the real-time formalism, where an $n$-point function at finite temperature can be seen as a weighted sum of an infinite number of $(n+2)$-point functions at zero temperature \cite{Kapusta:1989tk}. All the zero-temperature $(n+2)$-point functions are invariant under the R-symmetry. Hence, it is expected that the R-symmetry is preserved even at finite temperature. This holds true for any global symmetry of the theory generated by a bosonic charge.

 The preservation of the R-symmetry can also be understood holographically by considering the case of the duality between type IIB superstring on $ \text{AdS}_{5} \times S^5 $ and  $\mathcal N=4$ SYM in four dimensions. In this case, the R-symmetry algebra $\mathfrak{so}(6)$ is realized as the isometry algebra of $S^5$, and these isometries remain unaltered even when the non-compact component of the ten-dimensional spacetime is given by a black hole solution. This heuristically happens because the black hole solution does not affect the geometry of $S_5$. Therefore, the R-symmetry is preserved in the holographic dual description, which in the case of a black hole background corresponds to a finite temperature CFT.

  All the currents and the breaking terms corresponding to supersymmetry, superconformal, and R-symmetry are summarized in Table \ref{tab: Explicit can susy currents}.
  \begin{table}[]
    \centering
	\renewcommand{\arraystretch}{1.5}
	\begin{tabular}{|c|c|c|}
		\hline
		Charge   &Current:  $\omega_{a} J^{\mu}_{a}(x)$ & Breaking term: $\Gamma_a^\beta(\beta, \vec x)$ \\
		%\hline
		\hline
		$\op Q_\alpha^I$  & $\xi G^{\mu}(x)$ & $2 \, G_{\alpha}^{\, 0}(\beta,\vec x)$ \\
		\hline
		$\overline {\op Q}_{\dot \alpha, I}$ & $\xi \overline G^{\mu}(x)$ &  $2 \, \overline G_{\dot \alpha}^{\, 0}(\beta,\vec x)$\\ \hline 
        $\op S ^{\alpha}_{I}$ &  $\xi \sigma^{\nu} x_\nu \overline G^{\mu}(x)$ & $\left [(\beta-2 \tau)\left(\sigma^0\right)^{\alpha }+2(x_i-y_i)\left(\sigma^i\right)^{\alpha }\right ] \overline G^{\, 0}(\beta,\vec x) $ \\ \hline 
      $\op {\overline{S}}_{\dot \alpha}^{I}$  & $\xi \, \overline \sigma^{\nu } x_\nu G^{\mu}(x)$ & $\left [(\beta-2 \tau)\left(\overline \sigma^0\right)_{\dot \alpha }+2(x_i-y_i)\left(\overline \sigma^i\right)_{\dot \alpha}\right ] G^{\, 0}(\beta,\vec x) $ \\ \hline 
        $\op R^{IJ}$ & $J_R^{\mu , I J}(x)$ & $0$ \\ \hline 
		%\hline
	\end{tabular}
	\caption{\emph{Explicit currents and breaking terms for the symmetries of the global Superconformal Group. The $(\tau ,\vec y)$ coordinates locates the operator to which the breaking term is applied. To unclutter the notation, contracted spinor and R-symmetry indices have been left implicit.}}
	\label{tab: Explicit can susy currents}
	\end{table}
\subsection{$\mathrm U(1)_Y$ Bonus symmetry at finite temperature and non-renormalization in $\mathcal N = 4$ SYM}
 In $\mathcal N = 4$ SYM it is known that some two- and three-point functions satisfy a \textit{bonus} $\mathrm U(1)_Y$ symmetry \cite{Intriligator:1998ig,Intriligator:1999ff}. This symmetry is not exact since only a subsector is invariant and therefore it produces selection rules for $n$-point functions of short operators\footnote{The selection rule is also satisfied by the three-point functions of two short and one long operators.} only when $n = 2, 3$ in the planar limit. 
\newline It has been conjectured and argued in \cite{Brigante:2005bq} that the $\mathrm U(1)_Y$ bonus symmetry is preserved at finite temperature in the planar limit. By following the procedure adopted in Section \ref{sec: Broken Ward Identities at Finite Temperature}, we do not find any obstacles to this claim. In particular, the $\mathrm U(1)_Y$ bonus is a bosonic symmetry and it acts trivially on the spacetime coordinates. It is indeed a good opportunity to remind that continuous symmetries can be broken at a finite temperature only if the currents depend explicitly, i.e. not only through the fields, on the space-time coordinates (e.g. the case of the dilatation operator) or if the symmetry is generated by fermionic charges, which are anti-periodic on the thermal circle $S^1_{\beta}$ (e.g. supersymmetry). Nevertheless, the bonus $\mathrm U(1)_Y$ symmetry is rather special because it is not an exact symmetry of the theory and therefore the procedure in Section \ref{sec: Broken Ward Identities at Finite Temperature} cannot be applied. In particular the main obstacle is that it is not a symmetry of the action. In order to understand if this symmetry is preserved at finite temperature, it can be tested on the correlation functions between two short operators $\mathcal O_S$. In general, in the OPE between two short operators long operators $\mathcal{O}_{L}$ can appear as well,  therefore, in the OPE regime
\begin{equation} \label{eq: contr. lon}
    \left \langle \mathcal O_S(x) \mathcal O_S(0)\right \rangle_\beta = \ldots + \frac{f_{\mathcal O_S\mathcal O_S\mathcal O_L}b_{\mathcal O_L}}{2^{J}c_{\mathcal O_L}\beta^{\Delta_{\mathcal O_L}}}|x|^{\Delta_{\mathcal O_L}-2\Delta_{\mathcal O_S}} C_{J}^{(1)}\left(\frac{\tau}{|x|}\right)+\ldots \ ,
\end{equation}
 where we isolated the contribution of a long operator with a non-zero one-point function. The zero-temperature non-renormalization theorem tells us that the scaling dimensions $\Delta_{\mathcal{O}_S}$ are protected quantities on the conformal manifold. In the large $N$ limit, the coefficients $f_{\mathcal{O}_S \mathcal{O}_S \mathcal{O}_L}$ are also protected (as a consequence of the bonus symmetry). If the non-renormalization theorem hold at finite temperature, the equation \eqref{eq: contr. lon} would imply that either the scaling dimensions $\Delta_{\mathcal{O}_L}$ are protected (which is not true in general, since they are not protected at zero temperature), or $b_{\mathcal{O}_L} = 0$ for every long operator in the OPE between the two short operators $\mathcal{O}_S$. The latter condition is true in the strict large $N$ limit for the case of gauge invariant operators \cite{Brigante:2005bq,Furuuchi:2005qm}. However, as already noticed in \cite{Brigante:2005bq}, this is a large $N$ statement that does not hold in the finite $N$ case. Since it is not guaranteed that $b_{\mathcal{O}_L} = 0$ at finite $N$, the non-renormalization of the thermal two-point function $\left \langle \mathcal O_S(x) \mathcal O_S(0)\right \rangle_\beta$ does not hold in general.
 \newline On the other hand the $\mathrm U(1)_Y$ bonus symmetry is an approximate symmetry at large $N$. This suggests that the $\mathrm U(1)_Y$ bonus symmetry is indeed preserved at finite temperature (in the same approximation and in the same limit of the zero temperature case) because all the one-point functions that could spontaneously break the $\mathrm U(1)_Y$ are equal to zero at large $N$.
\section{(Super)Conformal Field theories on $T^2 \times \mathbb R^{d-2}$}\label{appendix:TorusCase}
\begin{table}[h!]
 \centering
 \small
	\renewcommand{\arraystretch}{1.5}
	\begin{tabular}{|c|c|c|}
		\hline
		Gen. &$\Gamma_a^0$ & $\Gamma_a^1$ \\
		\hline
		\hline
		$\op P_\mu$ & $0$ & $0$ \\
		\hline
		$\op D$ & $\beta T^{00}$ & $\beta' T^{11}$ \\
		\hline
		$\op L_{ij}$ & $0$ &  $0$\\
		\hline
		$\op L_{0i}$ &  $ \beta T^{0 i}$ & $0$ \\
        \hline
		$\op L_{1i}$ & $0$& $ \beta' T^{1 i}$ \\
        \hline
		$\op L_{01}$ &  $\beta T^{01}$ & $\beta' T^{01}$ \\
        \hline
        $\op K_0$ & $\beta \left[ (\beta-2 \tau_0) \, T^{00} +2 \tau_1 T^{01}+2  y_i T^{0i}\right]$&  $\beta' \left[ (\beta'-2 \tau_1) \, T^{01} +2 \tau_0 T^{01}+2  y_i T^{0i}\right]$
		\\
        \hline 
        $\op K_1$ &$\beta \left[ (\beta-2 \tau_0) \, T^{10} +2 \tau_1 T^{11}+2  y_i T^{1i}\right]$& $\beta' \left[ (\beta'-2 \tau_1) \, T^{11} +2 \tau_1 T^{11}+2  y_i T^{1i}\right]$
		\\ 
        \hline 
       $\op K_i$ & $\beta \left[ (\beta+2 \tau_0) \, T^{0j} +2  \tau_1 T^{1j}+2  y_i T^{ij}\right]$& $\beta' \left[ (\beta'+2 \tau_1) \, T^{1j} +2  \tau_0 T^{0j}+2  y_i T^{ij}\right]$ \\ 
		\hline
         $\op Q_\alpha$ &$2 \ G^{0}$ & $2\  G^{1}$\\ \hline
         $\overline{\op Q}_\alpha$ &$2 \ \overline G^{0}$ & $2 \  \overline G^{1}$\\ \hline
         $\overline {\op S}^{\dot \alpha}$ & $\left [(\beta-2  \tau_0)\left(\overline \sigma^0\right)+2\tau_1\left(\overline \sigma^0\right)+2y_i\left(\overline \sigma^i\right)\right ] G^{0}$ & $\left [(\beta'-2 \tau_1)\left(\overline \sigma^1\right)+2\tau_0\left(\overline \sigma^0\right)+2y_i\left(\overline \sigma^i\right)\right ] G^{1}$ \\ \hline
        $\overline {\op S}^{\dot \alpha}$ & $\left [(\beta-2\tau_0)\left(\overline \sigma^0\right)+2\tau_1\left(\overline \sigma^0\right)+2y_i\left(\overline \sigma^i\right)\right ] G^{0}$ & $\left [(\beta'-2 \tau_1)\left(\overline \sigma^1\right)+2\tau_0\left(\overline \sigma^0\right)+2y_i\left(\overline \sigma^i\right)\right ] G^{1}$ \\ \hline
        $\op R$ & 0 & 0  \\ \hline 
	\end{tabular}
	\caption{\normalsize\emph{Explicit breaking terms for the symmetries of the (Super) conformal Group on $T^2 \times \mathbb R^{d-2}$. Spinor and R-symmetry indices are suppressed for simplicity. The operators are computed in $(\beta,\tau_1,\vec x),(\tau_0,\beta',\vec x)$ depending if they appear in the $\Gamma_a^0,\Gamma_a^1$ column. We indicate with $\vec y$ the $\mathbb R^{d-2}$ distance between the insertion of the operator to which the current is applied and $(\tau_0/\beta,\beta'/\tau_1,\vec x)$ indicates the position of the operator. Modular invariance is encoded in the ``symmetry'' between the $\Gamma_a^0$ and the $\Gamma_a^1$ columns. }}
	\label{tab: Explicit T^2 breaking terms}
	\end{table}
We discuss here the breaking of the (Super) conformal Group on the manifold $T^2 \times \mathbb R^{d-2}$. We derive all the conformal and superconformal Ward identities for this geometry and we briefly discuss the immediate consequences. The procedure we adopt here is exactly the same of Section \ref{sec: Broken Ward Identities at Finite Temperature}, showing the generality of this procedure; for this reason, we skip some details, already discussed for the finite temperature case. The starting point is that on the geometry $\mathcal M_{\beta,\beta'} = T^{2}(\beta,\beta')\times \mathbb R^{d-2} =S_\beta^1\times S_{\beta'}^1\times \mathbb R^{d-2}$, the broken Ward identities take the form 
\begin{equation}
    i \omega_a \op G_a \left \langle \mathcal O_1(x_1) \ldots \mathcal O_n(x_n) \right \rangle_{\mathcal M_{\beta,\beta'}} =   \left \langle\delta S \ \mathcal O_1(x_1) \ldots \mathcal O_n(x_n) \right \rangle_{\mathcal M_{\beta,\beta'}} \ .
\end{equation}
The variation of the action is the integration of the derivative of the current over the whole manifold and we have therefore to consider \begin{equation}
    \begin{split}
       & \int_{T^2\times \mathbb R^{d-2}} d^d x \ \partial_\mu \left \langle J_a^{\mu} (x) \mathcal O_1(x_1) \ldots \mathcal O_n(x_n) \right \rangle_{\mathcal M_{\beta,\beta'}} = \\ & \ \ \ \ \  = \lim_{R\to \infty}\int_{T^2\times S_1^{d-3}} d \Omega \,  d \tau_0 \, d \tau_1 \ R^{d-3}  \left \langle J_a^{i} (\Omega,R,\tau_0,\tau_1) \mathcal O_1(x_1) \ldots \mathcal O_n(x_n) \right \rangle_{\mathcal M_{\beta,\beta'}}+\\ & \ \ \ \ \ \ \ \  \ \  \  + \int_{\mathbb R^{d-2}}d^{d-2} x \int_0^{\beta'} d \tau_1 \ \left \langle \Gamma_a^{0}(\tau_1,\vec x) \mathcal O_1(x_1) \ldots \mathcal O_n(x_n) \right \rangle_{\mathcal M_{\beta,\beta'}}+ \\ &  \ \ \ \ \ \  \ \ \ \ \   + \int_{\mathbb R^{d-2}}d^{d-2} x \int_0^\beta d \tau_0 \ \left \langle \Gamma_a^{1}(\tau_0,\vec x) \mathcal O_1(x_1) \ldots \mathcal O_n(x_n)\right  \rangle_{\mathcal M_{\beta,\beta'}} \  ,
    \end{split}
\end{equation}
where the first boundary term vanishes (or leads to a renormalized infrared divergence) because of clustering and the second term depends on the breaking terms \begin{equation}
     \Gamma_{a}^{0}(\tau_1,\vec x) = J^{0}(\beta,\tau_1,\vec x)- J^{0}(0,\tau_1,\vec x)\ , 
\end{equation}
\begin{equation}
    \Gamma_a^{1}(\tau_0,\vec x) = J^{1}(\tau_0,\beta',\vec x)- J^{1}(\tau_0,0,\vec x)\ .
\end{equation}
Similarly to the finite temperature case, these quantities can be explicitly computed and the results for the (Super) conformal Group are provided in Table \ref{tab: Explicit T^2 breaking terms}. 
 Also in the case of the torus, it is clear that translations, a subset of the rotations, R-symmetry, and any global symmetry are preserved. All the results appearing in the  table are in agreement with the expectations \cite{Luo:2022tqy}. In particular, modular invariance is explicit just by looking at Table \ref{tab: Explicit T^2 breaking terms}: when $\beta$ and $\beta'$ are exchanged,  the time direction (chosen to be $x^0$) and the direction $x^1$ can be interchanged as well to preserve the set of (broken) Ward identities.

 As in Subsection \ref{ssec: ham and mom}, we can write down Hamiltonian and momentum operators. However, in this case an interesting physical interpretation is missing. Nonetheless it is possible to write the implicit version of the Cardy formula as in the thermal case (cfr. equation \eqref{eq: Cardy formula}): \begin{multline}\label{eq: cardy torus}
     \left \langle T^{00}\right \rangle_{\mathcal M_{\beta,\beta'}}=  \frac{\beta}{d} \int d^{d-2}x \ \int_0^{\beta'} d \tau_1 \ \left \langle T^{00}(0,\tau_1,\vec x) T^{00}(0)\right  \rangle_{\mathcal M_{\beta, \beta'}} + \\ + \frac{\beta'}{d} \int d^{d-2}x \ \int_0^{\beta} d \tau_0 \ \left \langle T^{11}(\tau_0,0,\vec x) T^{00}(0)\right \rangle_{\mathcal M_{\beta, \beta'}}\ .
 \end{multline}
 The above equation corresponds to the dilatation broken Ward identity and it relates the free energy of the theory to the two-point function of the stress-energy tensor. It can also be generalized for any operator $\mathcal O$ \begin{multline}
     \Delta_{\mathcal O} \left \langle \mathcal O\right \rangle_{\mathcal M_{\beta,\beta'}}=  \beta \int d^{d-2}x \ \int_0^{\beta'} d \tau_1 \ \left \langle T^{00}(0,\tau_1,\vec x)\mathcal O(0) \right \rangle_{\mathcal M_{\beta, \beta'}} + \\ + \beta' \int d^{d-2}x \ \int_0^{\beta} d \tau_0 \ \left \langle T^{11}(\tau_0,0,\vec x) \mathcal O (0)\right \rangle_{\mathcal M_{\beta, \beta'}}\ .
 \end{multline}
 An interesting application of those equations concerns the deconstructions of six-dimensional theories (in particular the $\mathcal N = (2,0)$ theory) \cite{Arkani-Hamed:2001wsh, Hayling:2017cva, Niarchos:2019onf, Niarchos:2020nxk,Bourton:2020rfo,Giedt:2003xr, Kim:2018lfo}. Furthermore, one can observe that the dimensional analysis argument posed in Appendix \ref{Appendix1} can be easily adapted to the case of  $\mathcal M = T^2\times \mathbb R^{d-2}$. In particular the radii of both circles have to be taken into consideration and the dilatation broken Ward identity can be written as \begin{equation}\label{KallanT}
     \left(\op D +\beta \frac{\partial}{\partial \beta}+ \beta' \frac{\partial}{\partial \beta'}\right)\left \langle \mathcal O_1(x_1) \ldots \mathcal O_n(x_n) \right \rangle_{\mathcal M_{\beta,\beta'}} = 0 \ .
 \end{equation}

 \medskip
 
\section{Conclusions and future directions}
In this paper, the general procedure for computing constraints from broken symmetries on non-trivial manifolds is discussed, with a special focus on CFTs on the thermal manifold $\mathcal M_\beta = S_\beta^1\times \mathbb R^{d-1}$. The explicit relations between correlation functions are systematically derived, emphasizing that broken Ward identities typically relate an $n$-point function to an $(n+1)$-point function, see e.g. equations \eqref{eq: dilatation BWI}, \eqref{eq: rotation BWI}, \eqref{eq: 1pt functions dil}.\begin{comment} Intriguingly enough the dilatation broken Ward identity can be expressed as \begin{equation}
     \left (\op D + \beta \frac{\partial}{\partial \beta }\right)\left \langle \mathcal O_1(x_1) \ldots \mathcal O_n(x_n) \right \rangle_\beta = 0 \ .
 \end{equation}
 And for the case of $\mathcal M = T^2 \times \mathbb R^{d-2}$ is provided in equation \eqref{KallanT}.\end{comment}
This work is only the beginning of the exploration of the consequences of these novel equations.

The implications of these equations in the OPE regime of are explored, 
where they correctly reproduce known results established in previous works \cite{Iliesiu:2018fao,Benjamin:2023qsc}. This demonstrates the consistency of the derived broken Ward identities with existing knowledge in the field.
Furthermore, the paper highlights the precise connections between the broken Ward identities of the Conformal Group and the Hamiltonian (energy spectrum) and momentum at finite temperature, as well as the dilatation and boost operators at zero temperature. These relations provide novel insights into the behavior of the energy and the mass spectrum under the combined effects of temperature and conformal transformations. 

Using broken Ward identities we provide a clean derivation of an  implicit version of the Cardy formula in higher dimensions, expressed in equation \eqref{eq: Cardy formula}. We refer to this equation as implicit because it relates the free energy (or equivalently the one-point function of the stress-energy tensor) with an integrated two-point function of the stress-energy tensor. This equation has already appeared in \cite{El-Showk:2011yvt} and was derived using the observation that changing the temperature is equivalent to varying the zero-zero component of the metric.
Our derivation is a direct consequence of the breaking of dilatation invariance and it comes elegantly as a specific application of broken Ward identities. Moreover, broken Ward identities 
automatically 
allow for the generalization of
such a formula in the case of higher point functions and different operators (not necessarily the stress-energy tensor). 
Furthermore in \eqref{eq: cardy torus}  we provide a similar but novel implicit Cardy formula  for the geometry $\mathcal M = T^2 \times \mathbb R^{d-2}$. This equation, together with \eqref{eq: Cardy formula},  provides further constraints on any local CFT at finite temperature that go beyond the conformal bootstrap conditions \cite{El-Showk:2011yvt}.
Last week, a completely different approach, based on the thermal effective action, appeared in \cite{Benjamin:2023qsc} (see in particular equation (A.3)). In the latter case an explicit version of the Cardy formula for large dimension operators was derived.

In this paper we also extend our analysis to superconformal field theories at finite temperature, demonstrating that even though  superconformal symmetry is explicitly broken, the R-symmetry is preserved in any CFT at finite temperature. The special nature of the $\mathrm U(1)_Y$ bonus symmetry of $\mathcal N = 4$ SYM theory at zero temperature requires a different  discussion. Nonetheless we conclude that this symmetry is also preserved at finite temperature and in the large $N$ limit, as already anticipated in \cite{Brigante:2005bq}.

In addition our methods can easily be generalized to other manifolds. We explicitly discuss the case of a (super)conformal field theory on the manifold $\mathcal M = T^2\times \mathbb R^{d-1}$. The latter manifold has many interesting physical implications and it already appears in different contexts in literature (see for example \cite{Luo:2022tqy,Benjamin:2023qsc}). Furthermore, the torus compactification is relevant for deconstructing six-dimensional to four-dimensional superconformal field theories
\cite{Arkani-Hamed:2001wsh, Hayling:2017cva, Niarchos:2019onf, Niarchos:2020nxk,Bourton:2020rfo,Giedt:2003xr,Kim:2018lfo}.
Concerning this stand of research where CFTs are placed on more general curved backgrounds,  a generalisation of the embedding space formalism was proposed in \cite{Parisini:2022wkb} and more recently the ambient space
formalism \cite{Parisini:2023nbd}.
Even though our method is technically different from the one of \cite{Parisini:2022wkb,Parisini:2023nbd}, it would be interesting to compare results and maybe even combine them.

The question regarding the uniqueness of the solution of the bootstrap problem for thermal two-point functions is indeed an important one, and it has been previously discussed in the literature, including  \cite{Alday:2020eua}. The equations derived from Broken Ward identities in this paper provide additional non-trivial constraints for both zero and finite temperature CFTs, which go beyond the standard bootstrap constraints. They enrich the bootstrap problem and open up new avenues for studying and understanding thermal CFTs.

In Appendix \ref{sec: Exact computation of phiphi for a free scalar theory}, we present a detailed analysis of the exact computation of the two-point function of a free scalar theory in four dimensions. This is done by imposing only KMS invariance, the correct behavior in the Regge limit, and using the OPE expansion with the input of operator dimensions. While this method yields analytical results for the free scalar theory, difficulties arise when attempting to apply it to more complicated cases where the OPE expansion cannot be trivially factored into holomorphic functions.
However, it is important to note that the inability to factorize the OPE expansion in holomorphic functions does not necessarily imply that the two-point functions of more interesting theories cannot be computed using a similar approach. It is indeed possible that the problem turns out to be just of technical nature. The analysis in Appendix \ref{sec: Exact computation of phiphi for a free scalar theory} serves as a starting point and highlights the challenges that may arise in more general cases. Further investigations and developments are needed to explore the possibilities of applying this method to compute two-point functions in other theories of interest.

In yet a different strand of research it would be interesting to  combine
broken Ward identities with different (other than the OPE) expansions.
For example using the K\"allén – Lehmann representation which holds for general QFTs, we may be able to perturbe away from the vertical dashed red line in Fig. \ref{fig:Scheme}.

Assuming the eigenstate thermalization hypothesis (ETH) a finite temperature two-point function $\langle \phi(x)\phi(0)\rangle_\beta $ can be written as a four-point function at zero temperature $\langle \mathcal O_\Delta(\infty) \phi(x) \phi(0)\mathcal O_\Delta\rangle$, in the limit $x\to 0$, $\Delta \to \infty $, where $|x|\Delta$ is fixed \cite{Delacretaz:2022ojg,Delacretaz:2021ufg,Karlsson:2021duj,Karlsson:2019qfi,Fukushima:2023swb}. Inserting such an equation in the broken Ward identities, one-point functions of scalars can be computed as integrated four-point functions at finite temperature.

Finally, it is also possible to use broken Ward identities to derive exact constraints for non-local BPS observables such as Wilson loops in supersymmetric theories  generalizing 
 known zero-temperature results.  We are currently pursuing this direction of research.

 \acknowledgments
    It is a pleasure to thank 
       Ant\'onio Antunes, Davide Cassani, Alejandra Castro, Pietro Ferrero, Stefano Giusto, Apratim Kaviraj and Vasilis Niarchos for very useful discussions at different stages of the work.
    AM and EP have benefited from the German Research Foundation DFG under Germany’s Excellence Strategy – EXC 2121 Quantum Universe – 390833306.
    
\appendix 

\section{A comment on the dilatation broken Ward identity}\label{Appendix1}
The dilatation broken Ward identity plays a crucial role in this paper as it provides valuable insights into the theory and its embedding in theory space. While dilatation invariance is exact at critical points, such as in zero-temperature CFT, the violation of dilatation invariance carries important physical information about the theory. 
\newline By considering the thermal manifold $\mathcal M_\beta$ and the broken dilatation symmetry, we showed in Section \ref{sec: Broken Ward Identities at Finite Temperature} that the breaking term associated with the dilatation operator can be explicitly computed. This breaking term captures the thermal effects on the dilatation symmetry and provides a correction that accounts for the presence of temperature.
\newline 
We present here an alternative derivation of the dilatation broken Ward identities, 
shedding light on the interplay between dilatation invariance, thermal effects, and the underlying dynamics of the system. Let us start from the simple dimensional analysis of a generic thermal correlator  
\begin{equation}
    \left \langle \mathcal O_1(x_1) \ldots \mathcal O_n(x_n) \right \rangle_\beta = \frac{1}{\beta^{a_0}}\frac{1}{x_1^{a_1}}\ldots \frac{1}{x_n^{a_n}} f\left(\frac{x_1}{\beta},\ldots ,\frac{x_n}{\beta}\right) \ ,
\end{equation}
where $f$ is an unknown function. Dimensional analysis fixes \begin{equation}
    \sum_{i=0}^{n}a_i = \sum_{j = 1}^n \Delta_j  \ .
\end{equation} The above expression for the correlator can be further simplified by using $SO(d-1)$ invariance: however, this is not necessary for our purpose. We can simply observe that the correlation function is invariant under the transformations 
\begin{equation}
    \mathcal O_i \to \lambda^{-\Delta_i}\mathcal O_i \ , \hspace{1 cm} x_i \to \lambda_i x_i \ ,  \hspace{1 cm} \beta \to \lambda_i \beta \ .
\end{equation}
The first two transformations are simply dilatations, whereas the last one is a non-physical scaling of the temperature\footnote{This transformation is not physical because the temperature defines the theory in the theory space parameterized by the couplings and the temperature (cfr. Fig. \ref{fig:Scheme}).}. Nevertheless, the fact that the dilatation invariance is compensated by a scaling of the temperature implies that \begin{equation}
    \left (\op D + \beta \frac{\partial}{\partial \beta}\right) \left \langle \mathcal O_1(x_1) \ldots \mathcal O_n(x_n) \right \rangle_\beta = 0  \ .
\end{equation}
 Observe that the dilatation breaking term in Table \ref{tab:ExplicitThermalContributions} can be derived from this equation. To see this, we can consider that a scaling of the temperature is equivalent to a scaling of the metric component $g_{00}$, which is ``dual" to $T^{00}$ in the sense that the stress-energy tensor is the source of $g_{00}$ \cite{Papadodimas2010}. Let us consider the case of a one-point function for simplicity: then \begin{equation}
    \frac{\partial}{\partial \beta}\langle \mathcal O\rangle_\beta = - \frac{1}{\beta} \int d^{d-1}x\int_0^\beta d \tau  \ \langle T^{00}(\tau, \vec x) \mathcal O(0) \rangle_\beta = - \int d^{d-1}x \ \langle T^{00}(0,\vec x)\mathcal O(0) \rangle_\beta \ ,
\end{equation}
where the last step is justified by the fact that the stress-energy tensor component $T^{00}$ generates the Hamiltonian (and therefore it commutes with it). Finally \begin{equation}
    \Delta_{\mathcal O}\langle \mathcal O \rangle_\beta = \beta \int d^{d-1}x \ \langle T^{00}(0,\vec x) \mathcal O(0)\rangle_\beta \ .
\end{equation}
 \section{Explicit checks of broken Ward identities}\label{appendix A}
  \subsection{Consistency with the identity operator $\mathds{1}$}
    The simplest consistency condition for a broken Ward identity is its good behavior when a single copy of the identity operator $\mathds{1}$ is plugged into the equation. Let us consider the dilatation broken Ward identity. Due to the properties of the identity operator, the left-hand side of the dilatation broken Ward identity trivially identical to zero. Hence, the broken Ward identity reduces to 
    \begin{equation}
			 \beta \int d^{d-1}y \Braket{\Gamma^{\beta}_{\text{dil.}}(\vec y)}_{\beta}=0 \ . \label{eq: cons cond id}
    \end{equation}
    This specific case highlights the importance of the renormalization procedure, in particular, the renormalization of the IR divergence \eqref{eq: bdy IR div} 
    \begin{equation}
        \beta \int^{\text{ren.}} d^{d-1}y\Braket{\Gamma^{\beta}_{\text{dil.}}(\vec y)}_{\beta}=\beta \int d^{d-1}y \Braket{\Gamma^{\beta}_{\text{dil.}}}_{\beta}- \Braket{\Gamma^{\beta}_{\text{dil.}}}_{\beta} \times \text{Vol}\left[\cm_{\beta} \right] \ ,
    \end{equation}
    where we highlighted the difference between a renormalized integral (the one commonly used in the formulas of this work) and a non-renormalized one. Recalling that any thermal one-point function is constant, the one-point function of $\Gamma^{\beta}_{a}$ can be extracted from the first integral, making the whole expression identically equal to zero and proving the consistency condition \eqref{eq: cons cond id}.
    \subsection{Two dimensions and broken Ward identities}
    Another sanity check is given by the fact that in two dimensions  the dilatation broken Ward identity gives the correct expression of the one-point function of the stress-energy tensor. The broken Ward identities derived in this work cannot be directly checked, since in order to say something about an $n$-point thermal function, the information required includes a $(n+1)$-point thermal function. In higher dimensions, the number of points of the correlation functions has been reduced by studying the OPE regime. However, in two dimensions the equations can be checked straightforwardly since the thermal correlation functions are known. This comes from the equivalence between the two-dimensional cylinder and the thermal manifold, and the existence of a conformal map between the plane $\mathbb R^2\sim \mathbb C$, where all the functions are known, and the cylinder $\mathbb R\times S_\beta^1$. We recall that the conformal map is \begin{equation}
     \mathbb C \ni \omega = \sigma+i \tau =  \frac{\beta}{\pi} \ln z \ ,
 \end{equation}
 where we are using the convention $0 \le \tau \le \beta$ and $-\infty < \sigma <\infty$.  
\paragraph{Dilatation broken Ward identity} Let us test, as an example, the one-point function of the stress-energy tensor. The dilatation broken Ward identity reads \begin{equation}
     \Delta_{T} \left \langle T^{00}\right \rangle_\beta = \beta \int d y \ \left \langle T^{00}(0,y)T^{00}(0) \right \rangle_\beta \ . \label{eq: appA int}
 \end{equation}
 The stress-energy tensor on the complex plane can be written as $2 \pi T^{00}(z,\overline z) = T(z)+\overline T(\overline z)$, and it is known that \begin{equation}\label{eq:2dStressTensor}
    \left  \langle T(z) T(0) \right \rangle = \frac{c/2}{z^4} \ , \hspace{1 cm} \left  \langle \overline T(\overline z) \overline T(0) \right \rangle = \frac{c/2}{\overline z^4}\ .
 \end{equation}
 The application of the conformal map returns the two-point function on the cylinder  \cite{Datta:2019jeo,DiFrancesco:1997nk,Mussardo:2020rxh}\footnote{Here we are using the usual convention for two-dimensional CFT: in this convention $b_T>0$ for unitary theories.} \begin{equation}
     \left \langle T^{00}(\omega,\overline{\omega}) T^{00}(0,0) \right \rangle_\beta = \left(\frac{\pi}{\beta}\right)^4 \frac{c^2}{144 \pi^2}+\frac{c}{8 \pi^2}\left(\frac{\pi}{\beta}\right)^4 \left[\text{csch}^4\left( \frac{\pi \omega }{\beta }\right)+\text{csch}^4\left(\frac{\pi \overline{\omega} }{\beta }\right)\right]\
 \end{equation}
 which must be interpreted as the thermal two-point function to be inserted in the integral \eqref{eq: appA int}. Up to IR divergences renormalization, the integration returns 
 \begin{equation}\label{eq:1ptfunctionStress}
     \left \langle T^{00} \right \rangle_\beta = \frac{\pi}{3\beta^2} c \ ,
 \end{equation}
 which is correct in a free bosonic theory, i.e. when $c=1$ \cite{Iliesiu:2018fao}, but it is also the starting point to compute the Cardy formula in two dimensions \cite{Mukhametzhanov:2019pzy}. 
\newline Further observe that for a one-point function, it is trivial to check that the breaking term of the dilatation is given by $- \beta \partial_\beta$. For what is explained in Appendix \ref{Appendix1} this is already an explicit check of the broken Ward identities.
\newline \paragraph{Rotation broken Ward identity} The next check concerns the rotation broken Ward identity. In this case, we have that \begin{equation}
   -2 \left \langle T^{00}(0) \right \rangle_\beta =  i \left \langle \op S_{01} T^{01}(0) \right  \rangle_\beta = \beta \int d y \ \langle T^{01}(0,y) T^{01}(0)\rangle_\beta  \ . 
\end{equation}
All these correlators can be explicitly computed by using equation \eqref{eq:2dStressTensor}. Further observe that \begin{equation}
    \langle T^{01}(x) T^{01}(0) \rangle_\beta  = - \langle T^{00}(x) T^{00}(0)\rangle_\beta  \ ,
\end{equation}
and therefore the rotation broken Ward identity is satisfied with the same calculations we performed for the dilatation broken Ward identities. The rotation broken Ward identity can be reconducted to a dilatation broken Ward identity also in the case of the two-point function. This is not a generic statement, but it is a consequence of the chirality of the correlation functions (including the stress-tensor) in two dimensions. We do not expect this to be true also in higher dimensions.
\newline \paragraph{Special conformal transformations broken Ward identities:} The last check we have to perform regards the special conformal transformations. In this case, we have to check that \begin{equation}
    2 \tau \Delta_T \left \langle T^{00}(\tau,x) \right  \rangle_\beta  = (\beta-2 \tau) \beta \int d y \ \left \langle T^{00}(y) T^{00}(0) \right \rangle_\beta = \Delta_T (\beta-2 \tau) \left \langle T^{00}(\tau,x) \right \rangle_\beta \ ,
\end{equation}
which corresponds to the broken Ward identity for $\op K_0$; the broken Ward identity for $\op K_1$ follows naturally.
In this case we did not  have to compute anything explicitly, but we just applied the dilatation broken Ward identity. In order to see that the above equation is correct we have to consider that the one-point functions are invariant under time translations, therefore it is possible to compute them in $2\tau$, so that, calling $\tilde \tau = 2 \tau$ we can use the equivalence $\tau \equiv \tau+\beta$ to explicitly see that the broken Ward identity is just an identity\footnote{Observe that the periodicity of the time coordinates is a feature of the time coordinate of local operators in correlation functions. One may be tempted to conclude that every $\beta$ factor in the broken Ward identities is equivalent to zero, producing in this way some inconsistency; this is clearly not correct since $\beta$ is a fixed scale of the theory.}.
\newline When tested on the higher point functions, the simplest possible example is the case of the two-point function of scalar primaries of weights $(h, h)$, such that $2h = \Delta$; these two-point functions read \begin{equation}\label{eq:phiphi2d2ptfun}
    \left \langle \phi(\tau,\sigma) \phi(0,0) \right \rangle_\beta = \left(\frac{\pi}{\beta}\right)^\Delta  \operatorname{csch}^\Delta \left(\frac{\pi}{\beta}(\tau+i \sigma)\right) \operatorname{csch}^\Delta \left(\frac{\pi}{\beta}(\tau-i \sigma)\right) \ .
 \end{equation}
 Straightforward computations show that the breaking term of the Ward identity corresponding to $\op K_0$ is given by \begin{equation}
     \beta \tau \int d \tilde \sigma \  \left \langle T^{00}(0,\tilde \sigma )\phi(\tau,\sigma)\phi(0,0) \right \rangle_\beta + \beta \sigma  \int d \tilde \sigma \  \left \langle T^{01}(0,\tilde \sigma )\phi(\tau,\sigma)\phi(0,0)\right \rangle_\beta \ ,
 \end{equation}
 whereas the breaking term corresponding to $\op K_1$ is given by \begin{equation}
     \beta \sigma \int d \tilde \sigma \  \left \langle T^{00}(0,\tilde \sigma )\phi(\tau,\sigma)\phi(0,0)\right \rangle_\beta-\beta \tau  \int d \tilde \sigma \  \left \langle T^{01}(0,\tilde \sigma )\phi(\tau,\sigma)\phi(0,0)\right \rangle_\beta \ .
 \end{equation}
These breaking terms can be written as  differential operators acting on the two-point function in equation \eqref{eq:phiphi2d2ptfun}. However, in this way the broken Ward identities simply trivialize, showing that dilatations, rotations, and momentum (broken) Ward identities trivialize the special conformal transformation broken Ward identity.
\newline Similar checks can be also performed for the two-point function of the stress-energy tensor \eqref{eq:2dStressTensor}. All the results are in agreement with our expectations. 
\paragraph{Supersymmetry and superconformal broken Ward identities}
In order to test the supersymmetry broken Ward identities we can use the OPE \footnote{We focus on the analytic sector, but analogous equations can be written for the anti-analytic sector.} \begin{equation}
    G(z_1) G(z_2)  = \frac{2 c}{3 (z_1-z_2)^3}+ \frac{2}{z_1-z_2} T(z)+\ldots 
\end{equation}
to conclude that the superconformal descendent of the supercurrent is given by the stress-energy tensor $T$. This is clear also from the super-Virasoro algebra anti-commutator \begin{equation}
    \{G_n,G_m\} = 2 L_{n+m}+ \frac{c}{3}\left(n^2-\frac{1}{4}\right) \delta_{n+m,0} \ ,
\end{equation} 
remembering that \begin{equation}
    L_n = \oint \frac{dz}{2 \pi i} z^{n+1} T(z)  \ , \hspace{1 cm} G_m = \oint \frac{dz}{2 \pi i} z^{m+1/2} G(z)\ .
\end{equation}
We have therefore to check that \begin{equation}
    \left \langle T^{00} \right \rangle_\beta = 4\int d\sigma \ \left \langle G^{0}(\beta,\sigma) G^{0}(0,0) \right \rangle_\beta \ .
\end{equation}
It is easy to check that the above holds in any two-dimensional superconformal theory; in particular, the coefficient of the one-point function of the stress-energy tensor can be computed from the two-point function of the supercurrent and it is given by equation \eqref{eq:1ptfunctionStress}. 
   \section{Free scalar theory in $d = 4$} \label{sec: Free Scalar Tests of the Broken Ward Identities}
   We are going to test the broken Ward identities in the free scalar theory in four dimensions. In particular we will explicitly solve the bootstrap problem posed in Subsection \ref{subsec: Other consistency conditions} for the two-point function of two fundamental scalar fields $\phu(x)$. We will also consider the bootstrapped two-point function to extract several one-point functions, as it was done in \cite{Petkou:2021zhg,Iliesiu:2018fao}. Focusing on the one-point function of the stress-energy tensor $T^{\mu \nu}$, we will show that the relation \eqref{eq: a and b constraint} has a clear physical interpretation, providing a strong non-trivial check for the broken Ward identities.
    \paragraph{Computation of $\Braket{\phu(x)\phu(0)}_{\beta}$} \label{sec: Exact computation of phiphi for a free scalar theory}
    In Section \ref{subsec: constraints in the OPE regime} we computed the explicit expression of the two-point functions in the OPE regime directly from the dilatations broken Ward identity
    \begin{equation}
         f_\beta(\tau,r) = \sum_{\mathcal{O}\in \phu \times \phu}\frac{J!}{2^J (\nu)_J}\frac{1}{\beta^{\Delta_{\co}}}\frac{f_{\phu\phu\co} b_{\co}}{c_{\co}} \left(\tau^2+ r^2  \right)^{\frac{1}{2}\left(\Delta_{\co}-2\Delta_\varphi \right)} C_{J}^{(\nu)}\left(\frac{\tau}{\sqrt{\tau^2+r^2}}\right ) \ .
    \end{equation}
    Specializing the result for a scalar free theory, we can use the explicit OPE expansion \begin{equation}
		\phu(x)\times \phu(0) \sim 1+a \phu^2(0) + \underbrace{b_{\mu}\phu(0)\partial^{\mu}\phu(0)}_{\sim V^{\mu}(0)}+\underbrace{c_{\mu \nu}\phu(0)\partial^{\mu}\partial^{\nu}\phu(0)}_{\sim T^{\mu \nu}(0)}+ \dots \ , \label{eq: free scal 2pt}
	\end{equation}
	where the specific structure of the operators is not important for our purpose. The crucial information we can extract from the OPE is a formula for the conformal dimension of the operator $\mathcal{O} \in \phu \times \phu$ (excluding the identity operator $\mathds{1}$) as a function of its spin $J$. Recalling that $\Delta_{\phu}=\frac{d-2}{2}$, 
	\begin{equation}\label{eq: Dimensions of operator FST}
		\Delta_{\co}=d-2+J \ ,
	\end{equation}
	Using this formula, we can turn the sum over the operators in the equation \eqref{eq: free scal 2pt} into a sum over the spins $J$
		\begin{equation} \label{eq: thermal 2pts function}
			f_{\beta}(\tau, r)=\frac{1}{\left(\tau^2+r^2 \right)^{\frac{d-2}{2}}}+\sum_{J=0}^{\infty}k_{J}\frac{1}{\beta^{d-2+J}}\frac{J!}{2^{J}(\nu)_{J}} \left(\tau^2+ r^2  \right)^{\frac{J}{2}} C_{J}^{(\nu)}\left(\frac{\tau}{\sqrt{\tau^2+r^2}} \right) \ ,
		\end{equation}
  where we grouped all the dynamic information in the coefficients
	\begin{equation}\label{eq: kJ definition}
		k_{J}=\frac{f_{\phu\phu\co} b_{\co}}{c_{\co}} \ .
	\end{equation}
	Since the equation \eqref{eq: thermal 2pts function} is missing manifest periodicity over the thermal circle, it should be established by constraining the coefficients $k_{J}$. Since the coefficients $k_J$ are functions of the $b_{\mathcal{O}}$ one-point function data, imposing manifest periodicity to the two-point scalar function will provide information about the one-point functions of the scalar free theory at the same time \cite{Iliesiu:2018fao}. 
    \newline From now on, we will set $d=4$. Thanks to this choice, the equation \eqref{eq: thermal 2pts function} now reads
	\begin{equation} 
		f_{\beta}^{d=4}(\tau, r)=\frac{1}{\tau^2+r^2} +\sum_{J=0}^{\infty}\frac{1}{\beta^{2+J}}\frac{k_{J}}{2^{J}} \left(\tau^2+ r^2  \right)^{\frac{J}{2}} U_{J}\left(\frac{\tau}{\sqrt{\tau^2+r^2}} \right),
	\end{equation}
	where we used the identity 
	\begin{equation}
		C_{J}^{(1)}\left(x\right)=U_{J}\left(x \right), 
	\end{equation}
	with $U_{J}(x)$ being the $J^{\text{th}}$ Chebyshev polynomial of the second kind. These polynomials admit a closed formula
	\begin{equation}
		U_{J}\left(x \right)=\frac{\left(x+\sqrt{x^2-1} \right)^{J+1} -\left(x-\sqrt{x^2-1} \right)^{J+1}}{2\sqrt{x^2-1}}
	\end{equation}
	which can be plugged into the $d=4$ solution
	\begin{equation} 
		f_{\beta}^{d=4}(\tau, r)=\frac{1}{\tau^2+r^2} +\frac{1}{2 \beta^2 r}\sum_{J=0}^{\infty}k_{J}\left(\frac{i}{2 \beta} \right)^{J} \bigg[( r-i \tau)^{J+1}+(-1)^{J} (r+i\tau )^{J+1}\bigg] \ .
	\end{equation}
	The structure of the solution can be further improved by noticing that the flat space term can be rewritten as 
	\begin{equation}
		\frac{1}{\tau^2+r^2}=\frac{1}{2 \beta^2 r} \left(-\frac{1}{4} \right)\left(\frac{i }{2 \beta} \right)^{-2}\left[\left(r-i\tau \right)^{-1}+(-1)^{-2}\left(r+i\tau \right)^{-1}    \right]
	\end{equation}
	and by adding the following factor, identically equal to 0
	\begin{equation}
		0=\frac{1}{2 \beta^2 r} k_{-1}\left(\frac{i }{2 \beta} \right)^{-1}\left[\left(r-i\tau \right)^{0}+(-1)^{-1}\left(r+i\tau \right)^{0}    \right] \ .
	\end{equation}
	After having defined $k_{-2}=-1/4$, the solution reads
	\begin{equation}
		f_{\beta}^{d=4}(\tau, r)=\frac{1}{2 \beta^2 r}\sum_{J=-2}^{\infty}k_{J}\left(\frac{i}{2 \beta} \right)^{J} \bigg[( r-i \tau)^{J+1}+(-1)^{J} (r+i\tau )^{J+1}\bigg] \ .
	\end{equation}
  The expression can be reshaped by relabeling $J=\ell-2$ and imposing the parity $\tau\to - \tau$ to leave the two-point function invariant, forcing the sum to run only over even integers $\ell=2 n$: this can be achieved by setting $k_{\text{odd}}=0$
  \begin{equation}
      f_{\beta}^{d=4}(\tau, r)=\frac{\pi}{2 \beta r}\sum_{n=0}^{\infty}\frac{ k_{2n-2}}{\pi^{2n}}\left(-\frac{1}{4 }  \right)^{n-1} \left\lbrace\left[\frac{\pi}{\beta}( r-i \tau)\right]^{2n-1}+\left[\frac{\pi}{\beta}( r+i \tau)\right]^{2n-1}\right\rbrace \ . \label{eq: boot start pt}
  \end{equation}
    By introducing the dimensionless complex variables
    \begin{equation}
        w= \frac{\pi}{\beta} (r-i \tau )\ , \qquad \overline{w}=  \frac{\pi}{\beta} (r+i \tau) \ , \label{eq: w coord}
    \end{equation}
  we conclude that the two-point function can be reduced to
    \begin{equation}
        f_{\beta}^{d=4}(w ,\overline{w}) =\left( \frac{\pi}{\beta}\right)^2 \frac{g\left( w\right)+g\left(  \overline{w}\right)}{w+\overline{w}} \ , \label{eq: 2pt free}
    \end{equation}
    with the sums over $n$ converging for $0 < |w| <\pi$. $g(w)$ is an odd function of $w$, and similarly $g(\overline{w})$ is an odd function of $\overline{w}$. We can now solve the bootstrap problem posed in \ref{subsec: Other consistency conditions} for the function \eqref{eq: 2pt free}. The function $g(z)$ is periodic and it has a single pole in $z = 0$, plus all the poles that come from the periodicity of the function. The residue of the pole is fixed by $k_{-2} =- 1/4$ (and by periodicity, this is the residue of all the other poles as well).
    A function that satisfies all the analytic properties required is the hyperbolic cotangent  \begin{equation}
        g(w) = \coth \left(w\right)\ ,
    \end{equation}
    and therefore the candidate solution is given by \begin{equation}\label{eq:1}
        \widetilde f_{\beta}^{d=4}(w ,\overline{w}) =\left( \frac{\pi}{\beta}\right)^2 \frac{\coth \left(w\right)+\coth \left(\overline w\right)}{w+\overline{w}} \ . 
    \end{equation}
    Let us assume that another solution to the bootstrap problem exists: since the equation \eqref{eq:1} already encodes all the analytic properties that can be read from \eqref{eq: boot start pt}, then the new solution must have the following form
    \begin{equation}
         \widehat f_{\beta}^{d=4}(w ,\overline{w}) =\left( \frac{\pi}{\beta}\right)^2 \frac{\left(\coth \left(w\right)+\delta f(w)\right)+\left(\coth \left(\overline w\right)+\delta f(\overline{w})\right)}{w+\overline{w}} \ ,
    \end{equation}
    where $\delta f(w)$ is a holomorphic, regular, and periodic function. Thanks to its periodicity, $\delta f(w)$ (and similarly its complex conjugate) can be expanded in an infinite series of hyperbolic sines and cosines
    \begin{equation}
        \delta f(w)=\sum_{n} \bigg(a_{n} \sinh\left( nw \right)+b_{n} \cosh(n w) \bigg).
    \end{equation}
    We can read from the OPE \eqref{eq: boot start pt} that the solution must be odd in $w$, hence we set all the coefficients $b_n$ equal to 0 and we are left with
    \begin{equation}
         \widehat f_{\beta}^{d=4}(w ,\overline{w}) =\left( \frac{\pi}{\beta}\right)^2 \frac{\coth \left(w\right)+\coth \left(\overline w\right)}{w+\overline{w}}+\sum_{n}a_{n} \left( \frac{\sinh\left( nw \right)}{w + \overline{w}}+\frac{\sinh\left( n \overline{w} \right)}{w + \overline{w}}\right) \ . \label{eq: test regge}
    \end{equation}
    We can now test the solution candidate \eqref{eq: test regge} in the Regge limit. Recalling the Section \ref{subsec: Other consistency conditions}, we switch to Regge coordinates
    \begin{equation}
        w=\frac{\rho}{\eta - \sqrt{\eta^2-1}} \ , \quad \overline{w}=\rho \left(\eta - \sqrt{\eta^2-1}\right) \ ,
    \end{equation} 
    and the Regge limits of the terms composing the solution candidate \eqref{eq: test regge} are
    \begin{align}
        \frac{\coth \left(w\right)}{w+\overline{w}} &\xrightarrow{|\eta| \to \infty  } \frac{1}{2 \eta \rho} \ , & \frac{\coth \left(\overline{w}\right)}{w+\overline{w}} &\xrightarrow{|\eta| \to \infty  } \frac{1}{\rho^2} \ , \nonumber\\
        \frac{\sinh \left(n w\right)}{w+\overline{w}} &\xrightarrow{|\eta| \to \infty  } \frac{e^{2 n \eta \rho}}{4 \eta \rho} \ , & \frac{\sinh \left(n \overline{w}\right)}{w+\overline{w}} &\xrightarrow{|\eta| \to \infty  }\frac{n}{4 \eta^2} \ , 
    \end{align}
    where the check was performed for $n>0$\footnote{If $n<0$, we can reduce ourselves to the case considered knowing that $\sinh(-x)=-\sinh(x)$.} and a fixed value of $\rho$. We immediately notice that the corrections given by the hyperbolic sines are not allowed, since they are not polynomially bound in the Regge limit. We deduce that $\delta f(w)=0$ and $\widehat f_{\beta}^{d=4}(w ,\overline{w})=\widetilde f_{\beta}^{d=4}(w ,\overline{w})$. Therefore the two-point function of a free scalar theory reads \begin{equation}\label{eq: final two point function}
        f_\beta ^{d=4}(\tau,r) = \frac{\pi}{2 \beta r} \left \{\coth\left(\frac{\pi}{\beta}(r+i\tau)\right)+\coth\left(\frac{\pi}{\beta}(r-i\tau)\right)\right\}\ .
    \end{equation}
    The two-point function above satisfies all the bootstrap condition in \ref{subsec: Other consistency conditions} and agrees with the known result \cite{Rodriguez-Gomez:2021pfh}.
    \newline The procedure presented in this appendix can in principle be extended to other theories; however the possibility of reading a factorization of the form \eqref{eq: 2pt free} from the OPE is not guaranteed. In general, formulating an ansatz from the OPE can be very complicated, and we are not able to prove that the solution is unique. For instance, in the case of the thermal two-point functions in generalized free fields theory, the OPE reads 
    \small \begin{equation}
    \begin{split}
       f^{\text{GFF}}_\beta(\tau,r) =  \frac{1}{\left(\tau^2+r^2\right)^{\Delta_\varphi}} &+\frac{1}{2 r}\sum_{J = 0}^\infty \sum_{n = 0}^\infty  \frac{f_{\varphi \varphi (J,l)} b_{(J,n)}}{c_{(J,n)}}\left(\frac{1}{\beta}\right)^{2\Delta_\varphi+J+n}\left(\frac{i}{2}\right)^J(r-i\tau)^{J+\frac{n}{2}+1}(r+i\tau)^{\frac{n}{2}}+ \\ & +\frac{1}{2r} \sum_{J = 0}^\infty \sum_{n = 0}^\infty  \frac{f_{\varphi \varphi (J,l)} b_{(J,n)}}{c_{(J,n)}}\left(\frac{1}{\beta}\right)^{2\Delta_\varphi+J+n}\left(\frac{-i}{2}\right)^J (r+i\tau)^{J+\frac{n}{2}+1}(r-i\tau)^{\frac{n}{2}}\ .
       \end{split}
    \end{equation}
    \normalsize
    The complicated structure of the equation above makes it very difficult to guess a factorization similar to \eqref{eq: 2pt free}, and consequently, it is difficult to prove the unicity of the solution too.
    \paragraph{One-point functions from the two-point function}\label{subsec: one point functions}
    In the previous Section, we used the OPE regime to write the thermal two-point function as a function of complex coordinates with the general structure \eqref{eq: 2pt free}, then we explicitly solved the bootstrap problem for such function. We can now study the final solution \eqref{eq: final two point function} in the OPE regime again to extract all the one-point functions CFT data $b_J$, spin by spin. This procedure was already present in the literature on thermal one-point functions \cite{Iliesiu:2018fao,Rodriguez-Gomez:2021pfh,Petkou:2021zhg} and we were able to recover all the previously known results.
    \newline In order to extract the $b_J$ coefficients, we need to expand the equation \eqref{eq: final two point function} in the OPE regime. This can be achieved thanks to the series expansion of the hyperbolic cotangent
    \begin{equation}
        \coth(z)=\sum_{n=0}^{\infty} \frac{2^{2n} B_{2n}}{(2n)!} z^{2n-1} \ , \quad 0 < |z| < \pi \ , \label{eq: coth series}
    \end{equation}
    where $B_{2n}$ is the $2n^{\text{th}}$ Bernoulli number. Notice that we highlighted the domain of convergence: if we apply the equation \eqref{eq: coth series} to the solution \eqref{eq: final two point function}, we can immediately see that the domain of convergence corresponds to the OPE regime. The solution now reads 
    \begin{equation}
        f_\beta ^{d=4}(\tau,r) = \frac{\pi}{2 \beta r} \sum_{n=0}^{\infty} \frac{2^{2n} B_{2n}}{(2n)!}\left\lbrace\left[\frac{\pi}{\beta}( r-i \tau)\right]^{2n-1}+\left[\frac{\pi}{\beta}( r+i \tau)\right]^{2n-1}\right\rbrace  
    \end{equation}
    and can be compared to the equation \eqref{eq: boot start pt}. The two expressions coincide if 
    \begin{equation}
        \frac{2^{2n} B_{2n}}{(2n)!}=\frac{ k_{2n-2}}{\pi^{2n}}\left(-\frac{1}{4 }  \right)^{n-1} \ , \quad n \geq 0 \ ,
    \end{equation}
    which is consistent with the choice $k_{-2}=-\frac{1}{4}$ made in the previous Section. For $n \geq 1$, the Bernoulli numbers enjoy a relationship with the Riemann $\zeta$ function
    \begin{equation}
        B_{2n}=(-1)^{n+1}\frac{2 (2n)!}{(2 \pi)^{2n}}\zeta(2n) \ , 
    \end{equation}
    so that the expression above can be rewritten as
    \begin{equation}
        k_{2n-2}=2^{2n-1} \zeta(2n) \ .
    \end{equation}
    Recalling the change of variable $2n=\ell=J+2$ made in the Section \ref{sec: Exact computation of phiphi for a free scalar theory}, we can reintroduce the dependence on the spin $J$
    \begin{equation}
         k_{J}=\frac{f_{\phu \phu \co_{J}}b_{J}}{c_{\co_J}}=2^{J+1} \zeta(J+2) \ , \quad J \geq 2 \ , \quad  J \text{ even} \ .
    \end{equation}
   Recall that in the previous Section the condition $k_{\text{odd}}=0$ had already been imposed.
    This completely solves the bootstrap program for the thermal one-point functions of the free scalar theory in $d=4$
\begin{equation}
    b_{\mathds{1}}=1 \ , \quad b_{J}=0 \quad (J \text{ odd}) \ ,  \quad b_{J}=2^{J+1} \zeta(J+2) \frac{c_{\co_J}}{f_{\phu \phu \co_J}} \quad (J \text{ even}) \ . \label{eq: bj free}
\end{equation}
 
 \section{Broken Ward identities, OPE regime and Leibniz rule}\label{Susy in OPE}
A generic (broken) Ward identity in the OPE regime on the thermal manifold returns non-trivial information not only about the CFT at finite temperature, but also about the data of the CFT at zero temperature. In this Appendix, we study a relationship between the structure constants of the zero temperature CFT, making use of a generic broken Ward identity. We show that this recovers the known Leibniz rule at zero temperature. 
\newline
Let us consider the two-point function 
\begin{equation}
    \left \langle \mathcal O_1(x) \mathcal O_2(0) \right \rangle_\beta  \ .
\end{equation}
In the thermal OPE regime $|x|<\beta$, we can expand the two-point function as 
\begin{equation}
   \left \langle \mathcal O_1(x) \mathcal O_2(0) \right\rangle_\beta  = \sum_{\mathcal O \in \mathcal O_1\times \mathcal O_2} \frac{f_{\mathcal O\mathcal O_1\mathcal O_2}}{c_{\mathcal O}} |x|^{\Delta_{\mathcal O}-\Delta_{\mathcal O_1}-\Delta_{\mathcal O_2}-J} x_{\mu_1}\ldots x_{\mu_{J}}\left \langle \mathcal O^{\mu_1\ldots \mu_J}\right \rangle_\beta \ .
\end{equation}
To make the notation lighter, let us define  $\tau_{\mathcal O} = \Delta_{\mathcal O}-\Delta_{\mathcal O_1}-\Delta_{\mathcal O_2}-J$.
Let us assume that the CFT at zero temperature enjoys a continuous symmetry generated by the generator $\op G$ and that at finite temperature such symmetry is broken and encoded into a broken Ward identity. The left-hand side of the broken Ward identity will then look like 
\begin{equation}\label{eq: LHS non in OPE}
    \left \langle [\op G,\mathcal O_1] (x) \mathcal O_2(0)\right \rangle_\beta + \left  \langle \mathcal O_1(x) [\op G,\mathcal O_2] (0) \right \rangle_\beta \ ,
\end{equation}
while the right-hand side will be given by the breaking term, which can be expanded in the OPE regime in a fashion similar to the one adopted in Section \ref{subsec: constraints in the OPE regime} 
\begin{multline} \label{eq: leibniz intermediate}
        \int d^{d-1} y \ \left \langle \Gamma^{\beta}(\vec y)\mathcal O_1(x) \mathcal O_2(0)\right \rangle_\beta = \\  =   \sum_{\mathcal O \in \mathcal O_1\times \mathcal O_2} \frac{f_{\mathcal O\mathcal O_1\mathcal O_2}}{c_{\mathcal O}} |x|^{\tau_{\mathcal O}} x_{\mu_1}\ldots x_{\mu_{J}} \int d^{d-1}y\ \left  \langle \Gamma^\beta(\vec y)\mathcal O^{\mu_1\ldots \mu_J}(x)\right \rangle_\beta  \ .
\end{multline}
The broken Ward identity can be applied to the equation \eqref{eq: leibniz intermediate}, returning
\begin{equation} \label{eq: leib intint}
     \int d^{d-1} y \ \left \langle \Gamma^{\beta}(\vec y)\mathcal O_1(x) \mathcal O_2(0)\right \rangle_\beta =\sum_{\mathcal O \in \mathcal O_1\times \mathcal O_2} \frac{f_{\mathcal O\mathcal O_1\mathcal O_2}}{c_{\mathcal O}} |x|^{\tau_{\mathcal O}} x_{\mu_1}\ldots x_{\mu_{J}} \left \langle [\op G, \mathcal O^{\mu_1\ldots \mu_J}](x)\right \rangle_\beta 
\end{equation}
The equation \eqref{eq: LHS non in OPE} can be expanded in the OPE regime as well. Once combined with the equation \eqref{eq: leib intint}, it returns \begin{multline}\label{eq: Bootstrap susy one-point func.}
         \sum_{\widetilde{\mathcal O} \in [\op G,\mathcal O_1]\times  \mathcal O_2}
         \frac{f_{\widetilde {\mathcal O}[\op G,\mathcal O_1] \mathcal O_2}}{c_{\widetilde{\mathcal O}}} |x|^{\tau_{\widetilde{\mathcal O}}} x_{\mu_1}\ldots x_{\mu_{\widetilde{J}}}
         \left \langle \widetilde{\mathcal O}^{\mu_1\ldots \mu_{\widetilde{J}}} \right \rangle_{\beta} + \\  + \sum_{\widehat {\mathcal O} \in \mathcal O_1\times [\op G,\mathcal O_2]}\frac{f_{\widehat {\mathcal O}\mathcal O_1[\op G,\mathcal O_2]}}{c_{\widehat{\mathcal O}}} |x|^{\tau_{\widehat{\mathcal O}}} x_{\mu_1}\ldots x_{\mu_{\widehat{J}}}\left \langle \widehat {\mathcal O}^{\mu_1\ldots \mu_{\widehat{J}}}\right \rangle_\beta = \\   = \sum_{\mathcal O \in \mathcal O_1\times \mathcal O_2} \frac{f_{\mathcal O\mathcal O_1\mathcal O_2}}{c_{\mathcal O}} |x|^{\tau_{\mathcal O}} x_{\mu_1}\ldots x_{\mu_{J}} \left \langle [\op G, \mathcal O^{\mu_1\ldots \mu_J}] \right \rangle_\beta \ .
\end{multline}
In this equation, the left-hand side can be seen as a  zero temperature contribution, whereas the right-hand side corresponds to a finite temperature correction. The above equations relate one-point functions of different operators appearing in different OPEs and can be easily extended to the most general case of an $n$-point function. By defining $x = x_1-x_2$, we get \begin{multline}\label{eq: Bootstrap susy n-point func.}
         \sum_{\widetilde {\mathcal O} \in [\op G,\mathcal O_1]\times  \mathcal O_2}\frac{f_{\widetilde {\mathcal O}\mathcal O_1\mathcal O_2}}{c_{\widetilde{\mathcal O}}} |x|^{\tau_{\widetilde{\mathcal O}}} x_{\mu_1}\ldots x_{\mu_{\widetilde{J}}}\left \langle \widetilde{\mathcal O}^{\mu_1\ldots \mu_{\widetilde{J}}}(x_2)\mathcal O_3(x_3) \ldots \mathcal O_n(x_n)\right \rangle_\beta+ \\ + \sum_{\widehat {\mathcal O} \in \mathcal O_1\times [\op G,\mathcal O_2]}\frac{f_{\widehat {\mathcal O}\mathcal O_1\mathcal O_2}}{c_{\widehat{\mathcal O}}} |x|^{\tau_{\widehat{\mathcal O}}} x_{\mu_1}\ldots x_{\mu_{\widehat{J}}}\left \langle \widehat {\mathcal O}^{\mu_1\ldots \mu_{\widehat{J}}}(x_2) \mathcal O_3(x_3) \ldots \mathcal O_n(x_n)\right \rangle_\beta = \\ =  \sum_{\mathcal O \in \mathcal O_1\times \mathcal O_2} \frac{f_{\mathcal O\mathcal O_1\mathcal O_2}}{c_{\mathcal O}} |x|^{\tau_{\mathcal O}} x_{\mu_1}\ldots x_{\mu_{J}} \left \langle [\op G, \mathcal O^{\mu_1\ldots \mu_J}](x_2) \mathcal O_3(x_3) \ldots \mathcal O_n(x_n)\right \rangle_\beta  \ .
\end{multline}
The interpretation is the same as for the equation \eqref{eq: Bootstrap susy one-point func.}: the two terms on the right-hand side are the terms that appear at zero temperature, the right-hand side instead comes from the correction at finite temperature. Although these equations might look complicated, they can simply be interpreted as the Leibniz rule for the generator $\op G$
\begin{multline}\label{eq:LeibnizRule}
    [\op G, \mathcal O_1](x_1)\times \mathcal O_2(x_2)+\mathcal O_1(x_1)\times [\op G, \mathcal O_2](x_2) = \\ = \sum_{\mathcal O \in \mathcal O_1 \times \mathcal O_2}
f_{\mathcal O \mathcal O_1 \mathcal O_2} |x|^{\tau_{\mathcal O}}x_{\mu_1}\ldots x_{\mu_J} [\op G,\mathcal O^{\mu_1\ldots \mu_J}](x) \ .
\end{multline}
The above equations do not return any constraints at finite temperature. However, if we project both sides of the equation \eqref{eq: Bootstrap susy one-point func.} on the operator $[\op G, \mathcal O]$, we obtain the zero temperature relations between structure constants \begin{equation}
    \frac{f_{[\op G,\mathcal O_1] \mathcal O_2 [\op G,\mathcal O]}}{c_{[\op G, \mathcal O]}}+\frac{f_{\mathcal O_1[\op G,\mathcal O_2][\op G,\mathcal O]}}{c_{[\op G, \mathcal O]}} = \frac{f_{\mathcal O_1 \mathcal O_2 \mathcal O}}{c_{ \mathcal O}}\ .
\end{equation}
The latter can be simply obtained at zero temperature by taking \eqref{eq:LeibnizRule}, inserting the operator $[\op G,\mathcal O]$, and considering the correlators thereof. 
\newline The above equation is true for any CFT and for any generator. For instance, if $\op G = \op P_\mu$, we would obtain the relation between the primary and the conformal descendants thereof. If $\op G$ is a supercharge, then the above equation relates the structure constants of the superconformal primary with the structure constants of the super-descendants. The above equation is correct also in the case of dilatations and rotations: however, it is more interesting to think of their action on the two-point function as the action of differential operators, as it was done in Section \ref{subsec: constraints in the OPE regime}.

\newpage
  %%%%%Bibliography%%%%%
	 \bibliographystyle{JHEP}
	 \bibliography{Bwi}
	 %%%%%%%%%%%%%%%%%%%%%%
\end{document}